\documentclass[]{emulateapj}
\usepackage{amsmath}
\usepackage{graphicx}

\shorttitle{Stellar Velocity Dispersion of Inner Halo}
\shortauthors{King et al.}

\begin{document}

\title{Stellar Velocity Dispersion and Anisotropy of the Milky Way Inner Halo}

\author{Charles King III\altaffilmark{1}, Warren R.\ Brown, Margaret J.\ Geller, \and Scott J.\ Kenyon}
\email{cking@cfa.harvard.edu, wbrown@cfa.harvard.edu}   

\begin{abstract}

We measure the three components of velocity dispersion, $\sigma _{R},\sigma _{\theta 
},\sigma _{\phi }$, for stars within $6<R<30$ kpc of the Milky Way using a new 
radial velocity sample from the MMT telescope.  We combine our measurements with 
previously published data so that we can more finely sample the stellar halo. We use 
a maximum likelihood statistical method for estimating mean velocities, dispersions, 
and covariances assuming only that velocities are normally distributed.  The 
alignment of the velocity ellipsoid is consistent with a spherically symmetric 
gravitational potential. From the spherical Jeans equation, the mass of the Milky 
Way is $M\left(R\le12\text{ kpc}\right) = 1.3\times 10^{11}$ M$_{\sun}$ 
with an uncertainty of 40\%.  We also find a region of discontinuity, 
$15\lesssim R\lesssim 25$ kpc, where the estimated velocity dispersions and 
anisotropies diverge from their anticipated values, confirming the break observed by 
others.  We argue that this break in anisotropy is physically explained by coherent 
stellar velocity structure in the halo, such as the Sgr stream.  To significantly 
improve our understanding of halo kinematics will require combining radial 
velocities with future {\it Gaia} proper motions.

\end{abstract}


\affil{Smithsonian Astrophysical Observatory, 60 Garden Street, Cambridge, MA 02138}

\altaffiltext{1}{Pleiades Consulting Group, Inc., P.O. Box 531, Lincoln, MA 01773}

\section{INTRODUCTION\label{sec: Intro}}

	The global kinematics of halo stars encodes important information about the 
mass, structure, and formation of the Milky Way.  The existence of coherent 
structures, such as the Sgr tidal stream \citep{ibata94, majewski03}, demonstrates 
that at least part of the Milky Way halo emerged from the accretion of smaller 
galaxies as expected in hierarchical galaxy formation \citep[e.g.,][]{searle78}. 
Theoretical simulations suggest that the inner $R\lesssim 20$ kpc region of the 
stellar halo should be dominated by stars formed \textit{in situ} whereas the outer 
region should be dominated by accreted stars on increasingly radial orbits 
\citep{bullock05, abadi06, johnston08, zolotov09, font11, mccarthy12, rashkov13}.

	Modern studies based on observational data from the Sloan Digital Sky Survey 
(SDSS) support a dual component stellar halo \citep{carollo07, carollo10, beers12}. 
In this view, the inner halo $R<15$ kpc exhibits a flattened distribution of stars 
on radial orbits with no net rotation contrasted with the outer halo $R>20$ kpc 
characterized by a spherical distribution of stars with net retrograde rotation. 
\citet{schonrich11, schonrich14}, however, argue that the dual halo results from 
observational errors and selection effects, and cannot be distinguished from a 
single halo full of substructure.

	Intriguingly, many other observers find a discontinuity in both the number 
density and the orbital properties of halo stars near $R \simeq 20$ kpc. Star counts 
of main sequence turn-off stars \citep{bell08, sesar11}, RR Lyrae stars 
\citep{watkins09, sesar13}, blue horizontal branch (BHB) stars \citep{deason11}, and 
K giants \citep{kafle14} all exhibit a break in the number density around $R=16-26$ 
kpc. Radial velocity surveys of BHB stars imply systematically radial orbits in the 
inner and outer halo \citep{deason11, deason12} but more tangential orbits in the 
region $15<R<25$ kpc \citep{kafle12, deason13}.  The anisotropy, $\beta$, that 
depends on the ratio of tangential and radial velocity dispersions, provides a 
useful means for quantifying systematic velocity changes in the break region.

	Several studies have exploited large samples of halo stars to measure 
the velocity dispersion and anisotropy profiles beyond $R>10$ kpc with 
varying results. \cite{sirko04} employed a sample of $1,170$ BHB stars 
selected from the SDSS Data Release 4 to measure the anisotropy $\beta = 0.1 
\pm 0.2$, a result consistent with isotropy, for stars with a median distance 
from the Galactic center of $R\sim 25$ kpc. \cite{deason12} analyzed $1,933$ 
BHB stars from SDSS Data Release 8 (DR8) with $16 < R < 48$ kpc to find a 
radially biased anisotropy of $\beta = 0.5_{-0.2}^{+0.08}$. Looking for 
evidence of a multi-component halo, \cite{kafle12} analyzed an SDSS DR8 
sample of about $4,500$ BHB stars and found the anisotropy was radially 
biased, $\beta = 0.5$, for $9 < R < 12$ kpc and $25 < R < 56$ kpc. To their 
astonishment, they discovered a sharp dip, $\beta \sim -1.2$, in the 
anisotropy parameter profile at $R \simeq 17$ kpc that they could not explain 
as arising either from halo substructures or from accretion.

	Stellar velocities also provide information about the gravitational 
potential and mass of the Galaxy. The spherical \citet{jeans15} equation 
provides a quantitative link between observations and the underlying 
gravitational potential \citep[e.g.,][]{binney08}. Several sophisticated 
Jean analyses have been performed in recent years, exploring the importance 
of density profiles, anisotropy assumptions, and potential models 
with SDSS observations \citep{deason12, kafle12, kafle14, loebman14}. 
	This paper explores the kinematics of stars with $6<R<30$ kpc using new
and existing data sets.

	We present a spectroscopic radial velocity survey of 6,174 faint
$18<r<21$ F-type stars obtained with the Hectospec spectrograph on the 6.5m
MMT telescope.  We target F-type stars because they are the densest luminous
stellar tracer at heliocentric distances of 12 to 20 kpc.  Padova tracks
predict that a 10 Gyr old 0.8 $M_{\odot}$ star with $\rm{[Fe/H]}=-1.7$ has an
absolute magnitude of $M_{r}=+4.5$ \citep{girardi02, marigo08, bressan12}, a
value consistent with globular cluster observations \citep{newby11}. Thus
$r=20$ and 21 mag F-type stars probe the halo at 12 and 20 kpc heliocentric
distances, respectively.  To validate our results and improve our statistics,
we also make use of 13,480 F-type stars culled from the SDSS.  Because the
SDSS F star sample is shallower than our Hectospec F star sample, we also use
3,330 BHB stars culled from the SDSS to better constrain the more distant
region $R>15$ kpc.

	In the next section, we describe the three data sets used in our analyses. 
Section 3 discusses the geometry of projecting the components of velocity on the 
line of sight and develops maximum likelihood statistical methods to infer the 
distribution of velocities from the observed line-of-sight velocities. Section 4 
describes the mean velocity, dispersion, covariance, and anisotropy profiles; the 
tilt of the velocity ellipsoid; and the mass of the Galaxy. In Section 5, we test 
the assumption of independent, normally distributed velocities and investigate the 
potential effect of the non-normal velocity distribution of the Sgr stream on our 
analysis. We summarize our conclusions in Section 6.

\begin{figure}[tbp]
 \includegraphics[height=3.5in, angle=270]{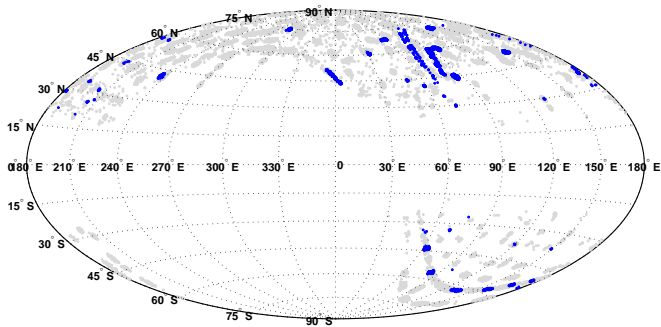}
	\caption{Angular distribution of F stars in the Hectospec (blue) and 
SDSS (gray) surveys in Galactic latitude and longitude.  The SDSS BHB 
stars have essentially the same footprint as the SDSS F stars (gray).}
\label{fig: Survey Maps}
\end{figure}

\begin{figure}[tbp]
\begin{center} 
 \includegraphics[width=2.25in, angle=270]{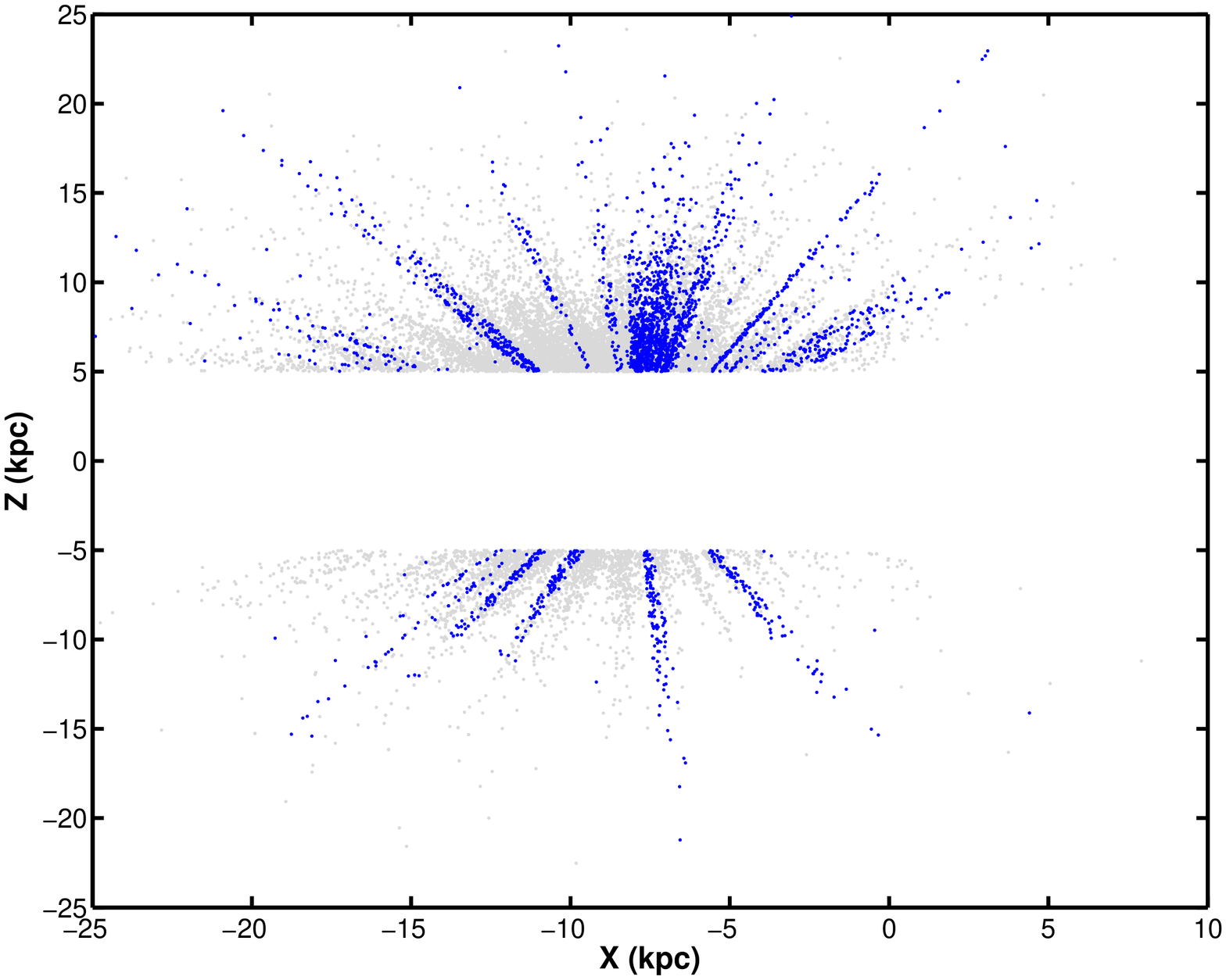} 
 \includegraphics[width=2.25in, angle=270]{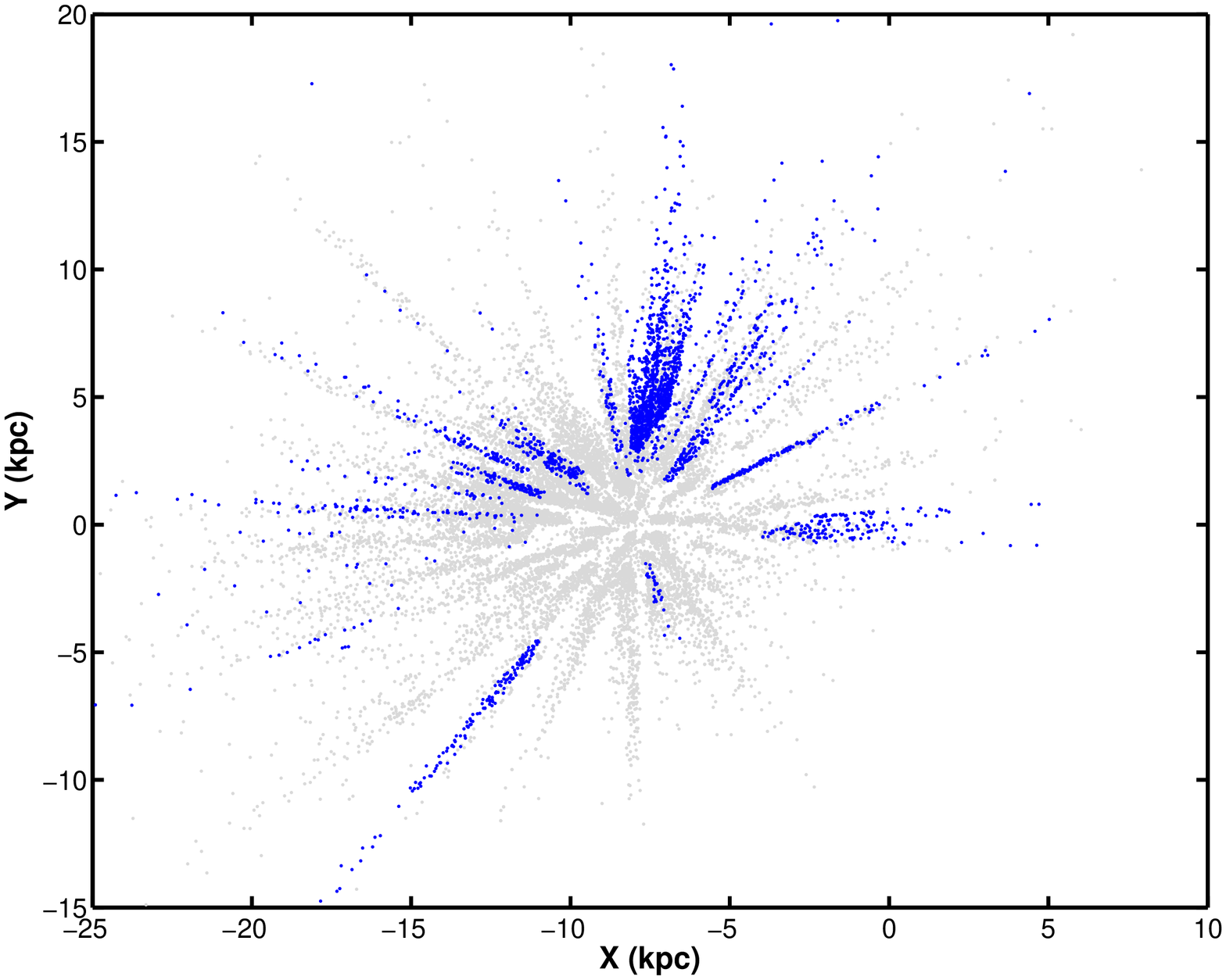}
\end{center} 
	\caption{Spatial distribution of F stars in the Hectospec (blue) and 
SDSS (gray) surveys projected onto the Galactic Cartesian coordinate X-Z and 
X-Y planes. The Sun is located at $X=-8$ kpc.}
\label{fig: XYZ}
\end{figure}

\section{DATA\label{sec: Data}}

	We analyze three data sets, including a new sample of radial 
velocities obtained with the Hectospec spectrograph on the 6.5m MMT 
telescope. To validate our results and improve our statistics, we add F-type 
stars culled from the SDSS Stellar Parameter Pipeline \citep[SSPP,][]{lee08, 
allende08} and the BHB star sample of \citet{xue11}.

\subsection{Hectospec F Star Sample\label{subsec: Hectospec}}

\subsubsection{Observations}

	Hectospec is a 300 fiber, multi-object spectrograph with a 1 degree
diameter field of view on the 6.5m MMT telescope \citep{fabricant05}. All
observations are made with the 270 line mm$^{-1}$ grating, which provides a
spectral resolution of 5 \AA\ over the spectral range 3700 -- 9100 \AA .  
Hectospec fibers are assigned such that high priority targets are assigned first,
followed by lower priority targets.  All targets and their priorities are
determined on the basis of SDSS de-reddened \citep{schlegel98} point spread
function magnitudes and colors.  We indicate de-reddened magnitudes and
colors with a subscript 0.

	We acquired 3,197 spectra between 2004 April and 2005 July, the first
year of Hectospec operations, with a dedicated halo star observing program.  Our
dedicated program targeted $17<g_0<20$ mag stars, prioritizing A-type (candidate
BHB) stars and filling the remaining Hectospec fibers with increasingly red stars
out to $(g-r)_0<0.5$. Given the relative surface densities of stars, 80\% of the
targets were F-type stars.

	We acquired an additional 8,143 spectra between 2009 January
and 2015 July with a parallel Hectospec observing program. Our parallel 
program took advantage of all unassigned fibers in Smithsonian Astrophysical 
Observatory (SAO) Hectospec programs, and filled those fibers with 
$18<r_{0}<21$ F-type stars. Our target selection for the parallel program 
prioritized the faintest stars at the main sequence turn-off $(g-r)_{0}\simeq 
0.25$. The parallel program yielded between 200 and 2,000 spectra a year, 
depending on fiber availability and Hectospec usage.

	Given the nature of our parallel observations, the overall angular 
distribution of F stars is irregular and clumpy, as seen in Figures \ref{fig: 
Survey Maps} and \ref{fig: XYZ}. The inner stellar halo of the Milky Way is 
expected to be spatially well-mixed, however \citep{bullock05}. Thus the F 
stars, selected by color, should be fair probes of halo kinematics.

\subsubsection{Radial Velocities}

	All Hectospec spectra are processed by the SAO Telescope Data
Center's data reduction pipeline \citep{mink07}. We visually inspect each
spectrum for quality control.  The major contaminants are M-type stars (4\%
of the spectra) and quasars and miscellaneous galaxies (1\% of the spectra).  
All contaminants are removed from the final Hectospec sample.

	We measure stellar radial velocities with the RVSAO cross correlation 
package \citep{kurtz98}. We use cross-correlation templates constructed from 
Hectospec observations of 33 bright A- and F-type stars with known velocities 
from the Century Survey Galactic Halo Project \citep{brown03, brown05b, 
brown08b}. We adopt the weighted mean velocity for objects with more than one 
observation (7\% of the spectra). Our median radial velocity uncertainty is 
$\pm15$ km~s$^{-1}$ for $r_0=20$ mag F-type stars.

	A cross-identification search finds 566 spectra in common with the SDSS SSPP 
catalog.  The mean difference between our radial velocity and SSPP 
\textit{elodiervfinal} is $1.0 \pm 20.4$ km~s$^{-1}$. Our systematic error with 
respect to SDSS is thus 1 km~s$^{-1}$, which is consistent with the sum of the 
measurement uncertainties.

\begin{figure}[tbp]
 \plotone{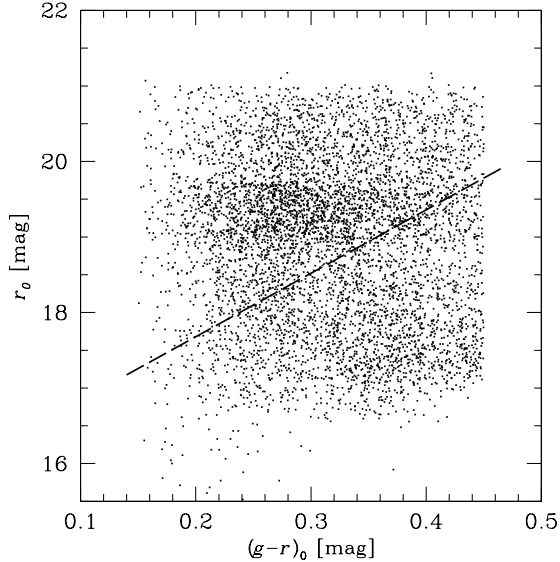}
	\caption{Color-magnitude distribution of the Hectospec sample of 
6,174 F-type stars.  All photometry comes from SDSS de-reddened point spread 
function magnitudes.  The median depth is $r_0=19.0$, or 8 kpc. Applying a 
$|Z|>5$ kpc cut removes the stars approximately below the dashed line.}
	\label{fig: TargetSelection}
\end{figure}

\subsubsection{Final Sample}

	Our targets were selected from a variety of SDSS data releases. We 
unify the photometry for this paper using SDSS Data Release 10 de-reddened 
point spread function magnitudes and colors.  Figure \ref{fig: 
TargetSelection} presents a color-magnitude plot of the cleaned Hectospec 
sample, which has a median depth of $r_0=19.0$ mag.

	Our spectra were acquired with a variety of exposure times and observing 
conditions. We therefore remove low quality observations as the first step of making 
a clean sample.  We require that each spectrum has more than $60$ instrumental 
counts and a signal-to-noise ratio per pixel $(S/N)>4$ in the continuum at 5000 \AA 
. Radial velocity measurements are possible from lower quality spectra, but our S/N 
requirement ensures repeatable velocities with $\leq $40 km~s$^{-1}$ errors. Second, 
we require that the ratio of counts in the continuum at 7800~\AA\ and 5000~\AA\ is 
$<1.25$. This requirement, given the Hectospec grating blaze function, eliminates 
all non-stellar objects, such as quasars, and spurious red objects, such as M-type 
stars.  Finally, we impose $0.15<(g-r)_{0}<0.45$ and $0.4<(u-g)_{0}<1.4$ to yield a 
clean sample of 6,174 F-type stars.

	Given the low spectral resolution and modest S/N ratio of the
Hectospec spectra, we do not perform stellar atmosphere model fits.  Instead, we
use the \citet{ivezic08} photometric parallax relation in combination with the
\citet{bond10} metallicity calibration.  These relations use de-reddened
broadband $ugri$ colors to estimate stellar luminosity $M_{r}$ and metallicity
[Fe/H]. Our radial velocity cross-correlation templates span A through F
spectral types, and the best-matching template to each spectrum validates this
approach.

	The average [Fe/H] $=-1.4$ and $M_{r}=+4.5$ for our entire sample are 
consistent with expectations of halo F-type stars \citep{ivezic08, bond10}. 
At faint $r>19$ magnitudes and low S/N ratio, however, the scatter in our 
estimated [Fe/H] becomes implausibly large. Spectroscopic metallicity 
measurements show that F-type stars at these depths have $-3<{\rm [Fe/H]}<-1$ 
and mean [Fe/H] $=-1.6$ \citep{allende14}. To minimize the distance error for 
stars with poorly estimated [Fe/H], we assign [Fe/H] $=-1.6$ to stars with 
[Fe/H] $<-3$ or [Fe/H] $>-1$.  We expect that our distance estimates are 
precise to about 15\% \citep{ivezic08, bond10}.  Our measured heliocentric 
radial velocities and estimated distances are provided in the Appendix 
\ref{app: DataTable}.

	To remove all significant disk contamination, we impose $|Z|>5$ kpc. 
This cut is motivated by the observed distributions of velocity and 
metallicity for F stars, which show that the disk population becomes 
negligible around $|Z|=4-5$ kpc \citep{ivezic08, carollo10, bond10, 
fermani13}. Restricting the sample to stars with Galactic rest frame 
velocities $|v_{rf}| <500$ km~s$^{-1}$ to eliminate potential unbound 
hypervelocity stars \citep{brown05} does not remove any stars from the 
sample. Our final Hectospec sample contains 3,049 F-type stars.

\subsection{SDSS F Star Sample\label{subsec: SDSS}}

	The SSPP catalog provides spectroscopic measures of effective 
temperature, surface gravity, and metallicity for stars observed by SDSS 
\citep{lee08, allende08}. We construct an F star sample from the SSPP catalog 
by selecting stars that have a stellar classification of F and extinction 
corrected colors satisfying $0.4<(u-g)_{0}<1.4$, $0.15<(g-r)_{0}<0.45$, and 
$-0.2<( r-i)_{0}<0.6$. We estimate heliocentric distances using the 
\citet{ivezic08} photometric parallax relation for consistency with the 
Hectospec sample. The median depth of the resulting F star sample is 
$r_{0}=17.4$, or $d \simeq 4 $ kpc.

	Requiring $|v_{rf}| <500$ km~s$^{-1}$ to avoid potential contamination by 
hypervelocity stars removes five stars, all of which have large radial velocity 
errors.  Removing all objects in common with the clean Hectospec F-star sample 
and performing a cut $|Z|>5$ kpc to remove potential disk contaminants yields a 
sample of 13,480 stars. The median heliocentric radial velocity uncertainty is 
$\pm 11$ km~s$^{-1}$ for $r=20$ mag F stars.

	Figure \ref{fig: Survey Maps} compares the angular distribution of 
the Hectospec and SDSS F star samples on the sky.  The overall distribution 
reflects the SDSS imaging footprint, which covers a large range of Galactic 
longitude but a restricted range of Galactic latitude, convolved with the 
spectroscopic survey regions.  Figure \ref{fig: XYZ} displays the 
corresponding spatial distribution of these stars projected onto X-Y and X-Z 
planes in Galactic Cartesian coordinates.  The majority of F stars are within 
about 10 kpc of the Sun and densely probe out to $R\simeq20$ kpc 
Galactocentric distances.

\subsection{SDSS BHB Star Sample\label{subsec: BHB}}

	\citet{xue11} spectroscopically identify 4,985 stars in SDSS as 
luminous BHB stars. These objects are evolved, metal-poor halo stars with 
typical absolute magnitudes of $M_{g}\simeq +0.8$ mag, significantly more 
luminous than the F-type stars.  Because the BHB stars come from SDSS 
spectroscopic fields, the sky coverage is essentially identical to that of 
the SDSS F star sample (Figure \ref{fig: Survey Maps}).

	We estimate BHB absolute magnitudes and distances using the 
\citet{deason11} color-magnitude relation.  The median apparent magnitude of 
the BHB sample is $g_{0}=16.7$ mag, corresponding to a depth of 15 kpc. 
Imposing a restriction on velocity that $|v_{rf}| <500$ km~s$^{-1}$ to avoid 
potential contamination by hypervelocity stars does not eliminate any stars. 
Retaining only stars with $|Z|>5$ kpc to avoid disk contamination yields a 
sample of 3,330 stars. The median heliocentric radial velocity uncertainty is 
$\pm 9$ km~s$^{-1}$ for $g=19$ mag BHB stars.

\begin{figure}[tbp]
 \includegraphics[width=3.5in]{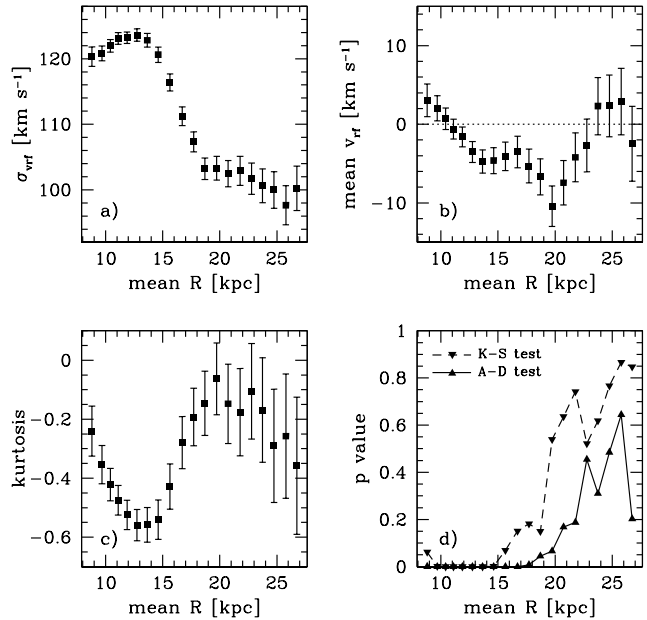}
	\caption{Statistics of the observed $v_{rf}$ distribution for the 
combined sample versus the mean $R$ for each bin.  Bins have fixed 4 
kpc width and thus overlap in $R$, with varying number of stars per bin. 
Panel a) is the velocity dispersion, panel b) is the mean velocity, panel c) 
is the kurtosis, and panel d) is the $p$ value of Kolmogorov-Smirnoff (K-S) 
and Anderson-Darling (A-D) tests for normality of the $v_{rf}$ distributions 
in each bin.} \label{fig: Obsvrf} \end{figure}

\subsection{Galactic Rest Frame Velocity\label{subsubsec: vrf_properties}}

	We transform all heliocentric velocities, $v_{helio}$, to 
Galactocentric rest frame velocities, $v_{rf}$, assuming a circular velocity 
of 235 km~s$^{-1}$ \citep{reid09, mcmillan10, bovy12, reid14} and a solar 
motion of $(U,V,W) = (11.1,12.24,7.5)$ km~s$^{-1}$ \citep{schonrich10},
 	\begin{align} {v_{rf}}& ={v_{helio}}+235\sin (l)\cos (b)+11.1\cos 
(l)\cos (b)  \notag \\ & +12.24\sin (l)\cos (b)+7.25\sin (b)\text{.} 
\end{align}

	Figure \ref{fig: Obsvrf} plots summary statistics for the 
observed $v_{rf}$ distribution for the combined data sets. We use bins of 
fixed 4 kpc width and plot points at the mean $R$ of each bin.  Figure 
\ref{fig: Obsvrf} (panel a) shows that the velocity dispersion of the 
combined data sets is relatively constant except for the range $15<R<18$ kpc, 
where the velocity dispersion declines suddenly from 123 km~s$^{-1}$ to 102 
km~s$^{-1}$.  Interestingly, the Hypervelocity Star survey measures a 
$113.9\pm6.6$ km~s$^{-1}$ velocity dispersion $15<R<20$ kpc \citep{brown10} 
in perfect agreement with the average velocity dispersion in this region.

	Figure \ref{fig: Obsvrf} (panel b) shows some apparent variation in the mean 
velocity of the stars, but for most bins the mean velocity does not differ 
significantly from 0 km~s$^{-1}$.  Figure \ref{fig: Obsvrf} (panel c) shows that the 
kurtosis of the distribution differs significantly from zero, however, indicating 
that the distribution of $v_{rf}$ departs from a normal distribution for bins $R<18$ 
kpc.  Both Kolmogorov-Smirnoff and Anderson-Darling tests significantly reject a 
normal distribution for bins $R<18$ kpc. We caution that both statistical tests are 
sensitive to the number of stars and note that the outer bins contain many fewer 
stars than the inner bins.

\section{THEORY AND METHODOLOGY \label{sec: TheoryAndMethodology}}

	There are few tests of the dynamical models used to derive Galactic 
parameters.  Halo models typically postulate a high degree of spatial 
symmetry and assume that the velocities of halo stars are normally 
distributed and uncorrelated with zero means.  The presence of structure and 
star streams expected in hierarchical galaxy formation might violate the 
standard assumptions.  To explore these ideas, we calculate the velocity 
means, dispersions, covariances, and the anisotropy of the inner halo from 
observational data using a minimal set of assumptions.

	Ideally, our input would be 3-dimensional velocities for a large number
of stars covering large areas of sky. Unfortunately, the typical $\pm 5$ mas
yr$^{-1}$ proper motion uncertainty \citep{monet03} of an $r=20$ mag, $d=12$ kpc
F-type star translates into a $\pm 284$ km~s$^{-1}$ uncertainty in tangential
velocity.  Tangential velocities are consequently uninformative except in large
statistical averages, which remain sensitive to systematic error
\citep{fermani13}. Radial velocities are about 25 times more accurate. The
typical $\pm 11$ km~s$^{-1}$ radial velocity uncertainty \citep{lee08} for an
$r=20$ mag star makes radial velocity our tool of choice.  Because the Sun lies
about $8$ kpc from the Galactic center, we use the heliocentric radial velocity,
combined with the angular position and the distance of each star, to constrain
Galactocentric tangential velocity components.

\begin{figure}[tbp]
 \plotone{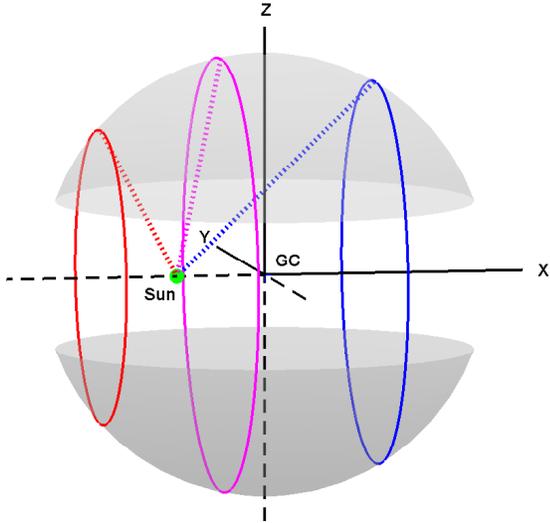}
	\caption{Viewing geometry depicting the relation between heliocentric and 
Galactocentric coordinates systems. Gray spherical caps show $R=20$ kpc for $|Z|>5$ 
kpc. Solid circles show where stars with heliocentric distances of $d=15$ (red), 20 
(purple), and 25 (blue) kpc fall on this $R=20$ kpc surface.}
	\label{fig: Geometry}
\end{figure}

\begin{figure*}[tbp]
 \plotone{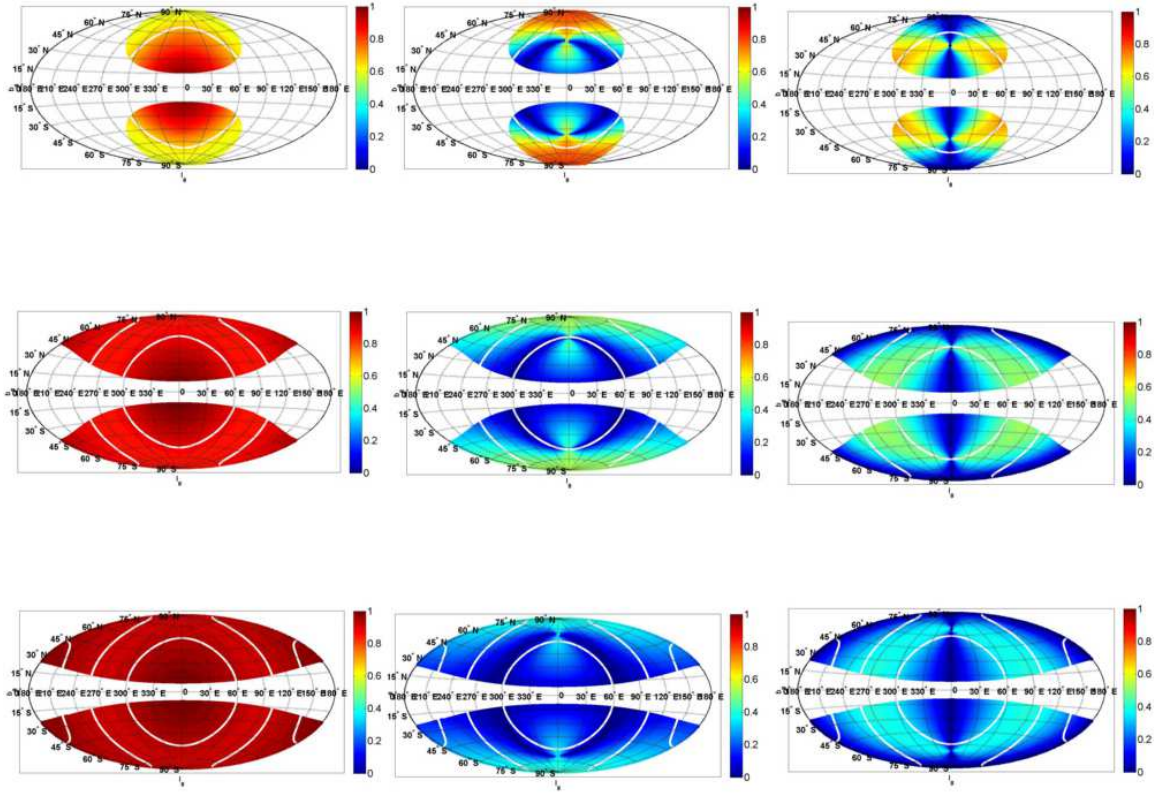}
	\caption{Line-of-sight projection factors at different Galactocentric 
distances in the halo. Columns show the components $p_{R}$, 
$p_{\protect\theta}$, and $p_{\protect\phi }$. Rows show the projection 
factor values at $R=10$, 15, and 20 kpc. For reference, white contours mark 
heliocentric distances of $d=15$, 20, and 25 kpc. } 
 \label{fig: Projection Factors} 
\end{figure*}

\subsection{Projection Factors\label{subsec: ProjectionFactors}}

	Geometrically, the observed line-of-sight velocity, $v_{los}$, is the 
projection of its Galactic velocity components, $\mathbf{v=} (v_{R},v_{\theta 
},v_{\phi })$, on the line-of-sight unit vector, $\mathbf{\hat{n}}$,
	\begin{equation}
v_{los}=\mathbf{v}\cdot \mathbf{\hat{n}=}\sum_{i=R,\theta,\phi}v_{i}p_{i},
	\label{eqn: vlos} \end{equation} where $p_{i}$ are projection factors 
for each component.  We use Galactic spherical coordinates because we are 
dealing with the halo.  The projection factors depend on angular position 
$(l,b)$, heliocentric distance $d$, and the distance from the Galactic center 
$R$ (see Appendix \ref{app: CoordinateSystems} for details).  Figure 
\ref{fig: Geometry} illustrates the geometry.  Sight lines from the Sun at 
different Galactic longitude, $l$, and latitude, $b$, intersect the gray 
$R=20$ kpc surface at points corresponding to different distances from the 
Sun, although they are all the same distance from the Galactic center.

	Importantly, different directions on the sky provide information about
different Galactic velocity components.  This point is widely understood but rarely
quantified.  Consider the region on the sky $|b|>80\arcdeg$:  70\% of $v_{los}$ is
in the $\theta$ component for stars at heliocentric distance $d=5$ kpc, but only
20\% of $v_{los}$ is in the $\theta$ component for stars at $d=15$ kpc; the
remainder is in the $R$ component.  Recovering tangential velocities from
$v_{los}$ clearly requires large observational samples with broad sky coverage.

	Figure \ref{fig: Projection Factors} quantifies the projection factors
from the observer's perspective as functions of angular position $(l,b)$ and
depth.  The three columns in Figure \ref{fig: Projection Factors} are for the
$p_{R}$, $p_{\theta }$, or $p_{\phi }$ projection factors.  Color indicates the
magnitude of the projection factor.  The white contours in each panel mark the
regions on the sky at heliocentric distances of $d = 15$, $20$, and $25$ kpc.  
The three rows in Figure \ref{fig: Projection Factors} represent Galactocentric
distances $R=10,15$, and $20$ kpc.  The sky coverage reflects the $|Z|>5$
restriction as $R$ increases.

	As expected, information about $v_{\theta }$ comes primarily from the
Galactic polar regions; information about $v_{\phi }$ comes primarily from
regions near Galactic latitudes of $60^{\circ }$ and $300^{\circ }$.  At large
distances, the observed line-of-sight velocity contains little information about
the tangential velocity components and becomes essentially radial. If there is
any correlation between angular position and kinematics, a systematic bias is
introduced.

\subsection{Maximum Likelihood Estimation\label{subsec: WtdLME}}

	We estimate the velocity means, dispersions, covariances, and anisotropy
of stars in the halo as functions of distance from the Galactic center using a
maximum likelihood procedure.  We adopt the standard assumption that the halo
star velocities have normal distributions \citep{binney08}; see Appendix
\ref{app: PDF} for details.  This minimal assumption allows us to test whether or
not the velocity components are normally distributed, have zero means, and are
uncorrelated.  Correlations among velocity components, if they exist, may reveal
underlying dynamic structures in the halo, such as star streams.

	Standard practice is to estimate the velocity dispersions by binning 
observations in a series of contiguous intervals in $R$. This procedure 
yields a series of discrete estimates, one for each bin.  We determine model 
parameters by maximizing the log likelihood function using analytic first and 
second derivatives and the R statistical software \citep{R14,maxlik12}.

	We validate our maximum likelihood calculations against a simulated 
set of stars with $6 < R < 30$ kpc and normal velocity distributions.  Our 
maximum likelihood calculation finds the correct solution for the underlying 
model parameters.

	We estimate both a restricted and a full statistical model. In the 
restricted model, the velocities are assumed to have zero means and to be 
uncorrelated, \textit{i.e.}, the covariance matrix is diagonal, as in 
previous studies of velocity dispersion in the halo. Estimating the full 
statistical model allows us to test empirically the assumptions of the 
restricted model that the velocities have zero means and are uncorrelated and 
thus test the fundamental assumptions underlying dynamical models of the 
Milky Way halo. For example, if the gravitational potential is spherically 
symmetric, the velocity-dispersion tensor is diagonal \citep{binney08}.

\section{RESULTS\label{sec: Results}}

	We begin by investigating the velocity dispersion and anisotropy of 
the Milky Way halo using our independent Hectospec F star sample.  We then 
combine our F star sample with the SDSS data sets to test our results with 
a larger set of observations.

\subsection{Anisotropy Parameter\label{subsec: Anisotropy}}

	One way to characterize the orbital structure of a spherical system, 
such as the halo, is through the anisotropy parameter defined as \citep[eq. 
4.61]{binney08}
	\begin{equation}
\beta =1-\frac{\sigma _{\theta}^2 +\sigma _{\phi}^2}{2\sigma_{R}^2}
\text{.}
	\end{equation} Because the anisotropy parameter depends on the ratio
of the velocity dispersions squared, it rapidly becomes more negative as the
tangential velocity dispersions increase relative to $\sigma_{R}$.  The value of
the anisotropy parameter is $\beta =1$ for perfectly radial orbits, $\beta =0$
for isotropy, and $\beta =-\infty$ for perfectly circular orbits.  From our
velocity dispersions estimates, we can quantify the anisotropy of the orbits in
the Milky Way halo and investigate the break in $\beta $ found in previous at
studies at $R\simeq20$ kpc \citep{kafle12, deason13}.

\begin{figure}[tbp]
 \includegraphics[width=3.5in]{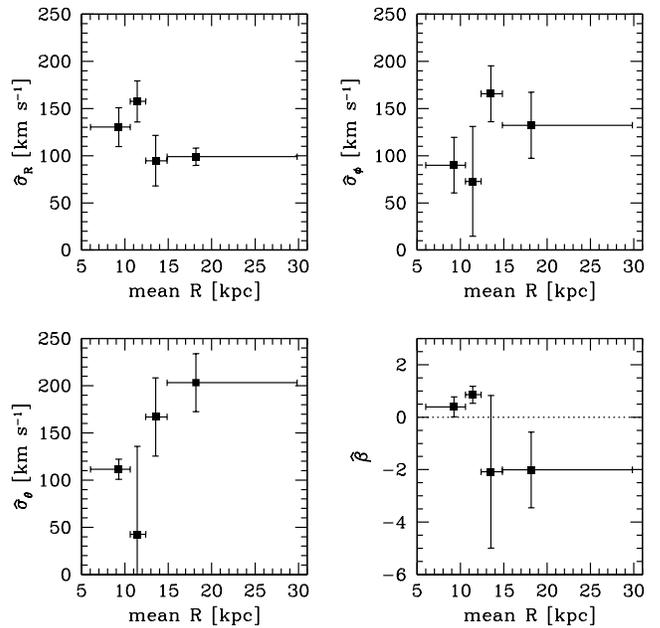}
	\caption{Velocity dispersions and anisotropy as function of $R$ for the 
Hectospec sample. The 3,049 stars in the Hectospec sample are partitioned into 
four radial bins of equal numbers. The estimates are plotted at the mean value of 
$R$ for stars in each bin.  Horizontal error bars show bin size; vertical 
error bars show $1\sigma$ uncertainties.  Appendix~\ref{app: ResultsTable}
presents the numbers. }
 	\label{fig: WRB F star results}
\end{figure}

\subsection{Results from Hectospec F Star Sample\label{subsec:FstarResults}}

	Using the line-of-sight velocities and positions from our Hectospec F
star sample, we explore the Galactic radial profile of the velocity dispersions
$\sigma _{R}, \sigma _{\theta }, \sigma _{\Phi}$, and the anisotropy parameter,
$\beta $.  We begin with the restricted model, which assumes that the velocity
components have zero means, $\mu _{i}=0$, and are uncorrelated, $\Sigma _{ij}=0$
for $i\neq j$. We relax both of these assumptions when we investigate the larger
combined data set in Section \ref{subsec: AllResults}.

\begin{figure}[tbp]
 \includegraphics[width=3.5in]{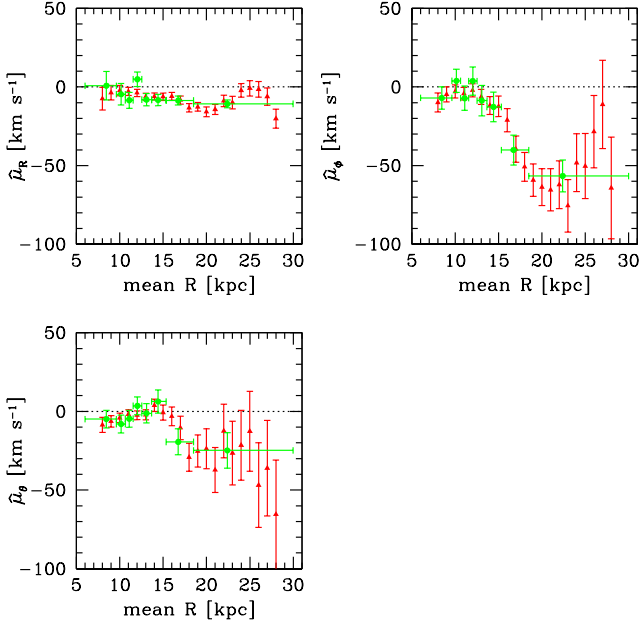}
	\caption{Mean velocity components of the combined sample.  The $19,859$ 
stars in the combined sample are partitioned into eight radial bins with equal 
numbers of stars (green circles) and also into overlapping bins of fixed 4 kpc 
width (red triangles).  Horizontal error bars show bin size; vertical error bars 
show $1\sigma$ uncertainties.  The estimate of $\protect\hat{\mu} _{\protect\phi }$ 
drops significantly between $15\lesssim R\lesssim 25$ kpc. }
	\label{fig: AllMeanV}
\end{figure}

\begin{figure}[tbp]
 \includegraphics[width=3.25in]{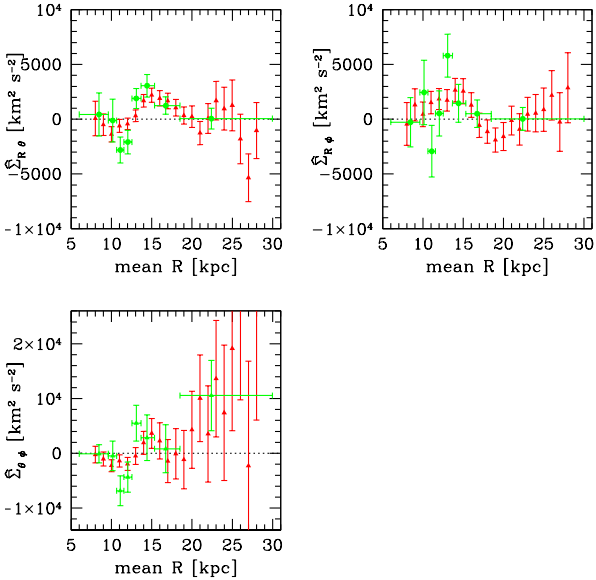}
	\caption{Velocity covariances for the combined sample.  Sample, error
bars, and bins as in Figure \ref{fig: AllMeanV}. Note the significant positive
covariance between $v_{R }$ and $v_{\theta}$ in the region $R = 14 - 17$ kpc. }
	\label{fig: AllCovar}
\end{figure}

\begin{figure}[tbp]
 \includegraphics[width=3.4in]{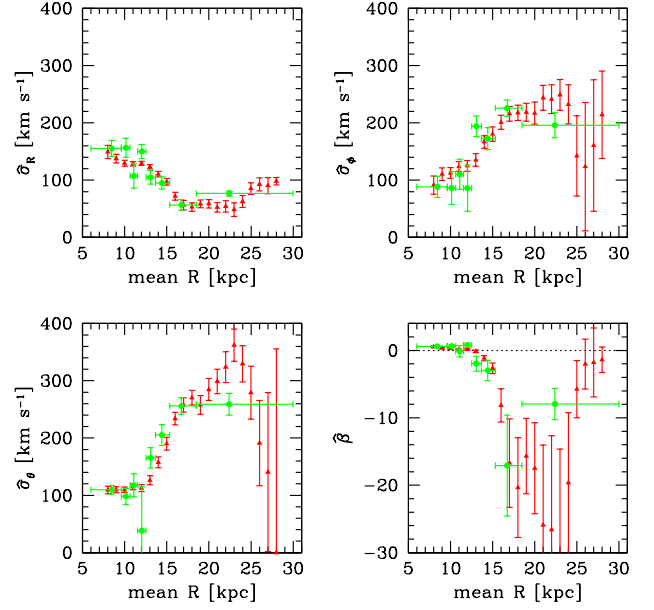}
	\caption{Velocity dispersions and anisotropy for the combined sample. 
Sample, error bars, and bins as in Figure \ref{fig: AllMeanV}.  The fixed width bins 
(red triangles) overlap in $R$ and have a varying number of stars per bin, whereas 
the fixed number bins (green circles) have varying widths that do not overlap.  
Either way, there is a large increase in estimated $\protect\hat{\sigma} 
_{\protect\theta }$ and $\protect\hat{\sigma} _{\protect\phi }$, and a corresponding 
drop in anisotropy $\protect\hat{\beta} $, in the range $15\lesssim R\lesssim 25$ 
kpc.}
	\label{fig: AllSigma}
\end{figure}

\begin{figure}[tbp]
 \includegraphics[width=3.1in]{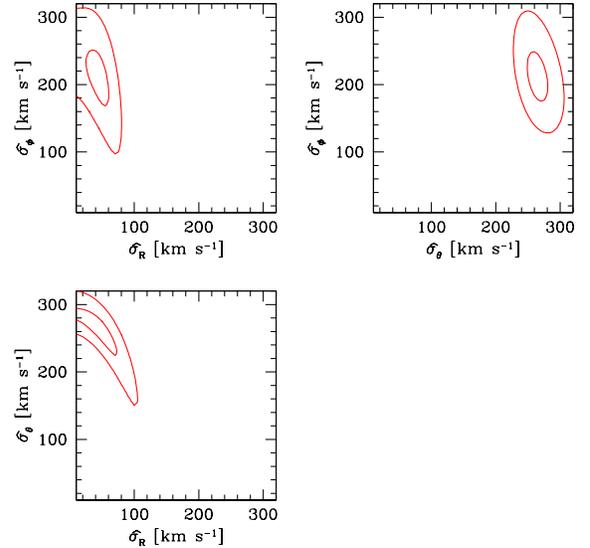}
	\caption{Contours of constant $\chi^2$ illustrating the correlation 
between the velocity dispersion components in the $14<R<18$ kpc bin. The dispersion
$\hat{\sigma}_{\phi}$ shows the largest range of values and thus is the most 
poorly constrained, while $\hat{\sigma}_{R}$ and $\hat{\sigma}_{\theta}$ are 
better constrained but correlated.}
	\label{fig: contours}
\end{figure}

	To estimate the parameters of our statistical model, we partition our
observations into four contiguous radial bins containing $n_{\rm bin} = $
762 or 763 stars and estimate the parameters and their errors separately for
each bin using maximum likelihood. A caret symbol distinguishes our parameter
estimates from the parameters themselves.  Figure~\ref{fig: WRB F star results}
and Appendix~\ref{app: ResultsTable} present our estimates of the velocity
dispersions and the anisotropy parameter. The solid points in Figure \ref{fig:
WRB F star results} represent the value of the point estimate for the
observations in a bin plotted at the mean $R$ for stars in that bin. The vertical
lines mark the $1\sigma $ confidence intervals for the estimates. Note that the
width of the bins varies as a function of Galactic radius, $R$, due to the
declining number density of stars and the limiting magnitude of the observations.
Taken together the graphs reveal how the velocity dispersions and the anisotropy
parameter vary with Galactic radius for the F stars in our sample lying more than
$5$ kpc above the Galactic plane with $6<R<30$ kpc.

	The results for $\hat{\sigma}_{R}$ and those for the first two points of 
$\Hat{\sigma} _{\theta },\hat{\sigma} _{\phi }$, and $\hat{\beta} $ are 
consistent with previous findings. The drop in the anisotropy parameter 
occurs near the discontinuity observed previously by others. Large 
values for the tangential velocity dispersions $\hat{\sigma} _{\theta }$ 
and $\hat{\sigma} _{\phi }$ lead to a negative value for $\hat{\beta}$ 
implying that orbits are tangentially biased at these radii. 

	These apparent anomalies could arise from several sources. The 
stellar velocities may not be normally distributed, or they may be correlated. 
The presence of structure or star streams in the inner halo may account for 
the discrepancies by violating the underlying assumptions of the spatial symmetry 
and statistical distribution of stellar velocities.  

\subsection{Results from Combined Sample\label{subsec: AllResults}}

	Combining the observations of the Hectospec and SDSS samples provides 
greater statistical power.  With the larger number of observations, we can 
explore a more complete model of the statistical distribution of halo stars.  
We relax the assumptions of the restricted model and include estimates of the 
means of the spatial velocities, $\mu _{i}$, to address the possibility of 
systematic motions, and the off-diagonal elements of the variance-covariance 
matrix, $\Sigma _{ij}$, to allow for correlated velocities.

	Figures \ref{fig: AllMeanV} - \ref{fig: AllSigma} and Appendix~\ref{app: 
ResultsTable} present the maximum likelihood estimates of the mean velocities, 
velocity dispersions, covariances among the velocities, and anisotropy parameter of 
the stars in the combined sample. The combined sample includes $19,859$ stars 
spanning $6 < R < 30$ kpc. We estimate the parameters as a function of Galactic 
radius using two methods for partitioning the data: 1) placing approximately equal 
numbers of stars ($n_{\rm bin} = $ 2,482 or 2,483)  within each interval in $R$ 
(green circles) and 2) using bins with a fixed width of $4$ kpc (red triangles). The 
error bars represent $1\sigma $. The two partition methods yield similar results.

	For $R<15$ kpc, the estimated mean velocities are small and, given the 
estimated errors and the uncertainties in the solar motion, are consistent with zero 
(Figure \ref{fig: AllMeanV}). Within the range $15\lesssim R\lesssim 25$ kpc, 
however, $\hat{\mu}_{\phi }$ and $\hat{\mu}_{\theta }$ drop significantly below 
zero.  Within $15\lesssim R\lesssim 25$ kpc, the stars have mean velocities 
$\hat{\mu}_{\phi } \simeq -50$ to $-80$ km~s$^{-1}$ and $\hat{\mu}_{\theta 
} \simeq -15$ to $-40$ km~s$^{-1}$.

	The estimated off-diagonal covariances among the velocity components (Figure 
\ref{fig: AllCovar}) are consistent with zero at the $2\sigma$ level with some 
exceptions. Positive covariances $\Sigma _{R \theta}$ between $v_{R }$ and 
$v_{\theta}$ and $\Sigma _{R \phi}$ between $v_{R }$ and $v_{\phi}$ occur in the 
region $R = 14 - 17$ kpc, for example.  The covariance estimates typically are 
at least an order of magnitude smaller than the variances, except for the covariance 
$\hat{\Sigma} _{\theta \phi}$ between $v_{\theta }$ and $v_{\phi }$ beyond $R 
\gtrsim 20$ kpc (Figure \ref{fig: AllCovar}), which is poorly estimated. The 
positive correlation between $v_{\theta }$ and $v_{\phi }$ occurs in approximately 
the same region where the mean velocity, $\mu_{\phi }$, turns significantly 
negative, again suggesting the possibility of correlated motions in this region.

	The velocity dispersion estimates exhibit unexpected behavior in the range 
$15\lesssim R\lesssim 25$ kpc. Over the entire span of $6 < R < 30$ kpc, the radial 
velocity dispersion, $\hat{\sigma} _{R}$, falls from about 150 km~s$^{-1}$ to a low 
of 50 km~s$^{-1}$ and then recovers to about 90 km~s$^{-1}$. The changes in the 
estimates $\hat{\sigma} _{\theta }$ and $\hat{\sigma} _{\phi }$ are more dramatic.  
Over the entire span, $\hat{\sigma} _{\theta }$ jumps from about 110 km~s$^{-1}$ to 
a maximum of about 360 km~s$^{-1}$; $\hat{\sigma} _{\phi }$ increases from about 110 
km~s$^{-1}$ to a peak of about 250 km~s$^{-1}$. Similarly, the anisotropy estimate, 
$\hat{\beta} $, declines from $0.5$ to around $-20$.

	To illustrate the correlations among the velocity dispersion components,
Figure \ref{fig: contours} plots contours of constant $\chi^2$ for stars in the
$14<R<18$ kpc bin. The velocity dispersion estimate $\hat{\sigma}_{\phi}$ shows the
largest range of values and thus is the most poorly constrained of the three
velocity components. The dispersions $\hat{\sigma}_{R}$ and $\hat{\sigma}_{\theta}$
are better constrained but correlated. Numerically, the maximum likelihood
calculation allows the two tangential components to settle around 300 km~s$^{-1}$ as
$\hat{\sigma}_{R}$ goes to zero. This is not a physically plausible scenario.  The
high velocity dispersions for $\sigma_{\theta }$ and $\sigma_{\phi}$ and the
positive covariance in the range $14<R<18$ kpc imply that some of the stars must
have velocities exceeding the Galactic escape velocity.

\begin{figure}[tbp]
 \includegraphics[width=3.5in]{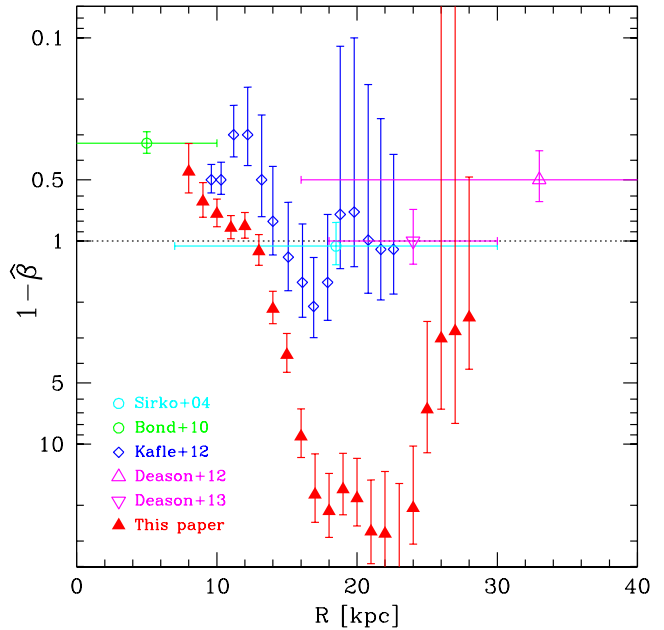}
	\caption{Comparison of anisotropy results.  We plot the logarithm of 
$(1-\hat{\beta})$ to present a more balanced comparison of tangential and radial 
anisotropies. We present anisotropy estimates for our overlapping fixed $4$ kpc bins 
(red squares), together with the anisotropy estimates of prior researchers.}
	\label{fig: Beta Comparison}
\end{figure}

	We compare our results for the anisotropy with previous research in 
Figure \ref{fig: Beta Comparison}.  We plot our $\hat{\beta}$ estimated in 
fixed width bins of $4$ kpc.  Our results at $R < 15$ kpc and at $R > 25$ kpc 
are in good agreement with prior research \citep{sirko04, bond10, kafle12, 
deason12, deason13}. In the range $15\lesssim R\lesssim 25$ kpc, however, our 
estimates of $\beta$ are substantially more negative (corresponding to 
tangentially biased orbits).

	We attribute the less negative values of $\beta $ estimated by other 
researchers to their use of much wider bins in $R$, which smooth the actual 
dispersion profiles; to a lower $Z$ cutoff of 4 kpc, which may increase 
contamination by disk stars; and to different statistical models that 
marginalize over tangential velocities and, in some cases, also impose additional 
assumptions about the gravitational potential and number density of stars. As 
shown by the horizontal bars in $R,$ the estimates of $\beta $ by 
\citet{sirko04}, \citet{deason12}, \citet{kafle12}, and \citet{deason13} are 
based on broad ranges in $R$ that extend beyond the interval where our results 
diverge. Although \citet{kafle12} do not report the sizes of the bins used for 
their estimations, we deduce from their original data that their bins increase in 
width from approximately $5$ kpc at $R=17$ kpc to $7$ kpc at $R=23$ kpc.

	At first glance, our results corroborate the discontinuity in 
$\hat{\beta} $ observed around $R\simeq20$ kpc.  Our results suggest the 
presence of correlated stellar motions within the region of $15\lesssim 
R\lesssim 25$ kpc, perhaps resulting from star streams or other structure.  
The implausibly high velocity dispersions also suggest that there may be 
another explanation for the discrepancy, namely that our underlying 
statistical model and those of previous researchers may be misspecified. We 
investigate these possibilities further in Section 5 
after first considering the robustness of our results, assessing the 
alignment of the velocity ellipsoid in the halo, and estimating the mass of 
the Milky Way interior to $R=12$ kpc.

\subsection{Robustness of Results\label{subsec: Robustness}}

	Our results are subject to various sources of potential error. We test 
the robustness of our parameter estimates with a series of sensitivity analyses 
on the input data. Systematically overestimating distances, for example, may 
inflate estimates of the tangential velocities at larger distances causing larger 
estimates of the tangential velocity dispersions, $v_{\theta }$ and $v_{\phi }$, 
and negatively biasing estimates of the anisotropy $\beta$.  To investigate the 
effect of systematic errors in stellar distance estimates, we recompute our 
results with all distances increased and decreased by 20\%. To investigate the 
effects of our choice of the location of the Sun with respect to the Galactic 
center, we recompute our results for $X_{\sun}=-8.5$ kpc. To investigate our 
choice of the Sun's circular velocity, we recompute our results for circular 
velocities of 200 and 250 km~s$^{-1}$. To investigate our sensitivity to 
different parts of the sky, we cut our sample in half in longitude and in 
latitude.

	We also investigate different model specifications, and compare the 
results from the restricted and unrestricted model for each sample. 
Finally, the location and size of the bins used in the estimations may 
affect the results. We therefore re-estimate the model parameters using an 
Epanechnikov kernel centered at $R$ with finite support on a total 
bandwidth of 4 kpc chosen to match the size of fixed bins used. This 
method gives greater weight to stars near $R$, rather than weighting them 
uniformly as occurs with fixed bins, and provides continuous estimates 
of the parameters as functions of $R$.

	In every case, the resulting parameter estimates change by less than about 
one standard deviation from our original results with the exception of a systematic 
distance error, where the change in parameter estimates is less than two standard 
deviations.  We conclude that our results are numerically robust within the 
estimated errors.

\subsection{Alignment of the Halo Velocity Ellipsoid and the Gravitational
Potential of the Milky Way\label{subsec: Alignment}}

	The alignment of the velocity ellipsoid for halo stars with respect 
to the Galactic coordinate system provides a powerful probe of the 
gravitational potential of the Galaxy. The alignment can be described by the 
tilt angles, $\alpha _{ij}$, derived by \citet{smith09a},
	\begin{equation} \tan \left( 2\alpha _{ij}\right) =\frac{2\Sigma 
_{ij}}{\Sigma _{ii}-\Sigma _{jj}}\text{,}
	\end{equation} where $\Sigma _{ij}$ is the covariance between the 
velocity components $v_{i}$ and $v_{j}$. The tilt angles specify the 
orientation of the velocity ellipsoid as the angle between the $i$-axis and 
the major axis of the ellipse resulting from the projection of the three 
dimensional velocity ellipsoid onto the $ij$ -plane (see, \emph{e.g.}, 
\citet{binney98} and \citet{smith09a}). \cite{smith09a} showed that if the 
inner halo is in steady state and the velocity ellipsoid is everywhere 
aligned with the Galactic spherical coordinate system, then the gravitational 
potential must be spherically symmetric. This result holds for the velocity 
ellipsoid of any tracer population, whether its density distribution is 
oblate, prolate or triaxial.

\begin{deluxetable}{ccccc}
\tablewidth{0pt}
\tablecaption{Tilt Angles\label{table: Tilt Angles}}
\tablecolumns{5}
\tablehead{
  \colhead{$\bar{R}$} & \colhead{$N$} & \colhead{$\alpha _{R\theta }$}
  & \colhead{$\alpha _{R\phi }$} & \colhead{$\alpha_{\theta \phi }$} \\
  \colhead{(kpc)} & \colhead{} & \colhead{(deg)} 
  & \colhead{(deg)} & \colhead{(deg)}
}
	\startdata
16.7 & 2482 & $ -2.3_{-2.9}^{+2.9}$ & $-1.2_{-5.9}^{+5.9}$ & $ 6.5_{-55.1}^{+47.2}$ \\
22.4 & 2483 & $ -0.1_{-3.6}^{+3.6}$ & $-0.1_{-7.3}^{+7.3}$ & $36.3_{-55.6}^{+24.9}$ \\
	\enddata
 \end{deluxetable}

	We find no evidence of any clear tilt in our analysis of the combined sample
using the eight disjoint, approximately equally populated bins of 2,842
or 2,843 stars in the interval $6<R<30$ kpc. None of the estimated tilt angles
differs from zero at a $1 \sigma $\ statistical significance level. For $R<15$ kpc,
the tilt angles are poorly estimated: the $1\sigma $ confidence intervals for the
estimates of the tilt angles are of order of tens of degrees. Our most tightly
constrained estimates occur for the bins with mean $\bar{R}=16.7$ $\left( R\in
\left( 15.3,18.5\right) \right) $ kpc and $\bar{R}=22.4$ $\left( R\in \left(
18.5,30.0\right) \right) $ kpc. The corresponding estimates for the tilt angles with
$1\sigma $ errors are given in Table \ref{table: Tilt Angles}. None is significantly
different from zero even at $1\sigma $. The tilt angle $\alpha _{\theta \phi }$ is
less well constrained due primarily to the large uncertainty in estimated covariance
between the two components.

	Our findings are consistent with a spherically symmetric 
gravitational potential in the inner halo, as suggested by several recent 
studies \citep{smith09a,koposov10,agnello12}. Our results accord with those 
of \citet{smith09a}, who found tilt angles consistent with spherical symmetry 
for $\sim 1,500$ nearby halo subdwarf stars with heliocentric distances of 
$\lesssim 5$ kpc (and $R\lesssim 11$ kpc) along the $\sim 250$ deg$^{2}$ 
covered by SDSS Strip 82. Their work was limited both in distance and sky 
coverage.  Our sample, on the other hand, extends to $R=30$ kpc with 
substantially greater sky coverage, including high Galactic latitudes where 
any contribution from the Galactic disk is negligible.

\subsection{Measuring the Interior Mass of the Milky Way \label{subsec: MWMass}}

	Many studies have exploited the dynamics of halo stars to measure the 
mass of the Milky Way \citep{xue08,deason12,kafle12} by assuming dynamical 
equilibrium and fitting specific forms of the distribution function, 
postulated to depend only on two integrals of motion, the binding energy and 
the angular momentum. The mass distribution of the Milky Way is commonly 
measured using the steady-state Jeans equation for a spherical potential 
using a tracer population of stars \citep[eq. 4.37]{binney08}:
	\begin{equation}
M\left( {<R}\right) =\frac{{R{\sigma _{R}^{2}}}}{G}\left( {-\frac{{d\ln {
\rho _{tr}}}}{{d\ln R}}-\frac{{d\ln \sigma _{R}^{2}}}{{d\ln R}}-2\beta }
\right) .  \label{eq: Jeans equation}
	\end{equation}
The mass interior to a radius $R$ is a function of the anisotropy, $\beta $, and
the logarithmic gradients of the radial velocity dispersion, ${\sigma _{R}}$, 
and the density of tracers, ${\rho _{tr}}$. The density distribution of 
stars in the halo has been extensively studied  
\citep[\textit{e.g.,}][]{yanny00, chen01, newberg06, juric08, deason11} 
leading to a recent consensus on the density profile \citep{deason11, 
sesar11} for the inner halo, $R<27$ kpc,
	\begin{align}
\rho \left( {{R_{q}}}\right) & \propto {R_{q}}^{\alpha }\text{,} \\
{R_{q}}^{2}& ={X^{2}}+{Y^{2}}+\frac{{{Z^{2}}}}{{{q^{2}}}}\text{,}  \notag
	\end{align}
where the power law index is $\alpha =-2.3$ to $-2.6$ and the minor axis 
to major axis ratio is $q=0.6-0.7$.
	Recent simulations by \cite{wang15} evaluate potential biases in 
estimating the mass of the Milky Way using dynamical tracers. They find that 
although deviations from spherical symmetry are relatively unimportant, 
deviations from dynamical equilibrium can cause significant bias. To avoid any 
discontinuity in the break region, where the assumption of dynamic equilibrium 
may not hold, we calculate the mass within 12 kpc of the Galactic center, 
$M\left(R\le12\text{ kpc}\right) $, using the results from the 4 kpc wide 
bin centered on 12 kpc and the Jeans equation (\ref {eq: Jeans equation}). 
We find $M\left(R\le12\text{ kpc}\right) = 1.3\times 10^{11}$ 
M$_{\sun}$.

	We estimate that the mass uncertainty is 40\%, arising largely from 
the uncertainty in the ${d\ln \sigma _{R}^{2}/d\ln R}$ term in Equation \ref{eq: 
Jeans equation}. The estimated error in $\sigma _{R}^{2}$ is only 10\%, and 
reported values of ${d\ln {\rho _{tr}/}d\ln R}$ also vary by 10\%.  When we 
compare adjacent bins to $R=12$ kpc to estimate the scatter in ${d\ln \sigma 
_{R}^{2}/d\ln R}$, however, we find that this term varies by 35\%.  To compare 
with previous work, we also calculate the mass for the $R=25$ kpc bin.  We find 
$M\left(R\le 25\text{ kpc}\right) \simeq 2.6\times 10^{11}$ M$_{\sun}$, a result 
that agrees to within 20\% with \citet{kafle12}.

\section{DISCUSSION\label{sec: Discussion}}

	Motivated by the large tangential velocity dispersion estimates in 
the region $15\lesssim R\lesssim 25$ kpc, we examine whether the 
observational data are consistent with our one fundamental assumption:  that 
the three components of velocity, $\left( v_{R},v_{\Theta },v_{\Phi }\right) 
$, are normally distributed. The normal distribution is described by two 
parameters, its mean and dispersion.  If the distribution of halo star 
velocities is not intrinsically normal and we (incorrectly) try to represent 
it as such, we expect the mean will remain largely unchanged but the 
dispersion will vary.  Notably, the Sgr stream passes through our survey 
region in a way that introduces a bimodal velocity distribution of stars into 
the sample beginning at $R\sim15$ kpc.

\subsection{Departures from Normality\label{subsubsec: VrfNormality}}

	If the underlying theoretical assumptions are correct, the observed
line-of-sight velocities should be normally distributed since linear combinations
of normally distributed random variables are also normally distributed.  Yet
Section \ref{subsubsec: vrf_properties} shows that the distribution of observed
radial velocities departs from normal in certain regions and fails standard
statistical tests for normality.  Physically, known or unknown star streams or
other velocity structure may be responsible for this observed departure from
normality.  Correlations between angular position and velocity break our model
assumptions.

	To better understand the possible causes of these discrepancies, we look 
more closely at the break region by selecting the $4,077$ stars in the 
$14<R<18$ kpc bin and plotting the histogram of $v_{rf}$ in Figure \ref{fig: 
Vrf Histogram}. The histogram reveals deviations from a normal distribution 
both in the center and in the wings of the distribution. The 
Kolmogorov-Smirnov test rejects normality at a reasonable level of 
statistical significance ($p=0.069)$, and the more powerful Anderson-Darling 
test that gives more weight to the tails than the Kolmogorov-Smirnoff test 
strongly rejects normality ($p=9.9\times 10^{-8}$).

\begin{figure}[tbp] 
\begin{center} 
 \plotone{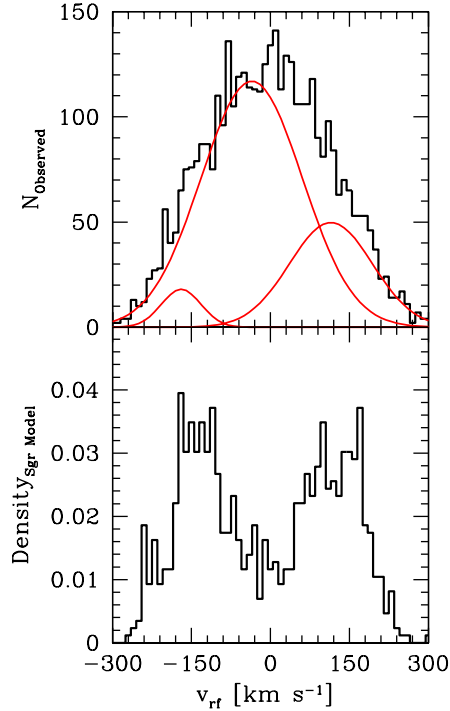} 
\end{center} 
     \caption{Top panel: Observed $v_{rf}$ distribution of stars in the bin
$14<R<18$ kpc, superimposed with a fitted, three-component, normal 
mixture model \citep{mixtools09}. There appears to be an excess of stars 
in both the negative and positive velocity wings of the distribution. 
Bottom panel: Distribution of the \citet{law10} Sgr $N$-body model 
sampled in the identical way for $14<R<18$ kpc.  The simulated Sgr stars
have a bi-modal distribution of line-of-sight velocities in this region. } 
\label{fig: Vrf Histogram}
\end{figure}

\subsection{Non-Normality of Sgr Stream Velocities\label
{subsubsec: Non-Normality of Sgr Velocities}}

	At least one significant halo structure lies within our survey 
footprint: the Sgr stream. To assess the impact of the Sgr stream, we turn to 
the $N$-body model of \citet{law10}. Their model is designed to match 
existing observational constraints on the location and motion of the Sgr 
stream in a triaxial potential.  We sample the $N$-body model for test 
particles that lie at $|Z|>5$ kpc in the range $14<R<18$ kpc and that fall 
within an approximation of our survey footprint on the sky ($b\geq 30^{\circ 
}$ or ($b\leq -30^{\circ }$ and $l\in \left( 50^{\circ},200^{\circ }\right) 
$)).  Although the majority of Sgr is at larger distances, a fraction of the 
stream is present in the $14<R<18$ kpc region.  Figure \ref{fig: Vrf 
Histogram} (lower panel) shows the $v_{rf}$ distribution of Sgr $N$-body 
particles in this region.

	Because the Sgr stream wraps around the sky in a roughly polar orbit, 
the Sgr stars passing through the $14<R<18$ kpc region exhibit multiple 
velocity peaks and a broad velocity dispersion. Suggestively, the two main 
peaks in $v_{rf}$ from the Sgr model coincide with the excesses found in the 
wings of the observed halo star distribution (Figure \ref{fig: Vrf 
Histogram}) in a fitted, three component, normal mixture model 
\citep{mixtools09}.

	Previous research estimates that up to $10\%$ of the stars in the survey 
region may originate in the Sgr stream \citep{king12}. The presence of Sgr stream 
stars in our sample may thus account, at least partially, for the observed 
discrepancies in the observed velocity dispersions and anisotropy of the halo in 
the break region.  We test this hypothesis by removing all stars in our sample 
within $10\arcdeg$ of the Sgr stream, $|B|<10\arcdeg$, using the Sgr stream 
coordinates, $(\Lambda, B)$, defined by \citet{belokurov14}.

	Removing the region of the Sgr stream from our sample modestly
lowers our estimates of the dispersions, which, in turn, increases our
estimate of the anisotropy.  The anisotropy in the $14<R<18$ kpc bin changes from
$\beta= -8$ to $\beta= -6$, in better agreement with previous results (Figure
\ref{fig:  Beta Comparison}). The change in $\beta$ is formally less than
$1\sigma$, but there may be other star streams for which we do not account.
Indeed, \citet{bell08} and \citet{schlaufman09} identify 30\% - 40\% of F-type
stars at 15 kpc depths in coherent spatial or velocity structures, and
\citet{janesh15} show that the percentage of halo stars in coherent structures
increases with depth. We conclude that the traditionally assumed normal velocity
distribution model may not properly represent the substructure of the stellar
halo.

\section{CONCLUSION\label{sec: Conclusion}}

	We use a large sample of 19,859 stars at $6<R<30$ kpc to investigate 
the mean velocities, velocity dispersions, covariances, and anisotropy of the 
Milky Way halo.  This dense sample enables finer binning than previously used 
and allows us to investigate the $15\lesssim R\lesssim 25$ kpc anomaly in 
anisotropy at higher statistical significance.

	We begin by presenting a new radial velocity survey of 6,174 faint 
F-type stars observed with the Hectospec spectrograph using the MMT 
telescope.  F-type stars are dense tracers of both the thick disk and halo.  
To focus on halo kinematics, we restrict our analysis to stars with $|Z|>5$ 
kpc.  We add stars from published SDSS radial velocity samples to create a 
combined sample of 19,859 stars that span $6<R<30$ kpc.

	We use the Sun's offset from the Galactic center to recover tangential 
velocity information from the observed line-of-sight velocities utilizing 
standard statistical methods. We make the minimal assumption that the underlying 
stellar velocity distribution is normal. We use a maximum likelihood procedure to 
calculate the velocity means, dispersions, covariances, and anisotropy. We find 
that the alignment of the velocity ellipsoid is consistent with a spherically 
symmetric gravitational potential. From the spherical Jeans equation, we estimate 
the mass of the Milky Way within 12 kpc is $M\left(R\le12\text{ 
kpc}\right) = 1.3\times 10^{11}$ M$_{\sun}$ with an uncertainty of 
40\%.

	A significant region of discontinuity $15\lesssim R\lesssim 25$ kpc
exists where the estimated velocity dispersions and anisotropy diverge from
their anticipated values, confirming the break region observed by others.  
The estimated tangential velocity dispersions in this region are so large
that stars would be unbound, an unphysical result.  Yet the results are
numerically robust. In sensitivity analyses (\textit{i.e.}, using a different
solar motion, different distance scale, different survey footprint, etc.),
the maximum likelihood calculation yields parameters that change by less than
about one standard deviation from our original result. Conversely, if we
input simulated data drawn from known normal velocity distributions, the
maximum likelihood estimation finds the correct velocity dispersion and
anisotropy parameters.

	We suggest that the discontinuity in the region $15\lesssim R\lesssim 
25$ kpc arises from the failure of the normal distribution model to describe 
the actual velocity data. Physically, known or unknown star streams or other 
velocity substructure may be responsible for the departure from normality. 
The predicted contribution of the (polar orbiting) Sgr stream in our survey 
region, for example, is a bi-modal distribution of stars in the wings of the 
observed radial velocity distribution. Sgr by itself cannot explain the 
discontinuity, but Sgr is unlikely to be the only structure in the halo.  

	The upshot is that larger radial velocity samples alone cannot improve
our understanding of the halo using the standard statistical approach.
Significant improvement requires direct tangential velocity constraints for halo
stars, like those soon to be provided by {\it Gaia}.  The combination of our
radial velocity measurements with {\it Gaia} proper motions will thus be very
useful for understanding the physical nature of the $15\lesssim R\lesssim 25$ kpc
discontinuity region and the kinematics of the Milky Way halo.

\acknowledgements

We thank Mike Alegria, Perry Berlind, Mike Calkins, Nelson Caldwell, Scott 
Gotilla, John McAfee, Erin Martin, and Ale Milone for their assistance with 
observations obtained at the MMT Observatory, a joint facility of the 
Smithsonian Institution and the University of Arizona.  We also thank William 
B.\ Simpson for helpful comments, Alis J.\ Deason for kindly providing data 
for Figure \ref{fig: Beta Comparison}, and David R.\ Law for kindly providing 
the Sgr $N$-body model.  CK dedicates this work to Carlos A.\ Berenstein. 
This paper uses data products created by the OIR Telescope Data Center, 
supported by the Smithsonian Astrophysical Observatory, and data products 
from the Sloan Digital Sky Survey, which is managed by the Astrophysical 
Research Consortium for the Participating Institutions. This research also 
makes use of NASA's Astrophysics Data System Bibliographic Services. This 
work was supported by the Smithsonian Institution.  CK gratefully 
acknowledges additional support from the Pleiades Consulting Group, Inc.

\textit{Facilities:} \facility{ \facility{MMT (Hectospec spectrograph)}}
\clearpage


\begin{thebibliography}{74}
\expandafter\ifx\csname natexlab\endcsname\relax\def\natexlab#1{#1}\fi

\bibitem[{{Abadi} {et~al.}(2006){Abadi}, {Navarro}, \& {Steinmetz}}]{abadi06}
{Abadi}, M.~G., {Navarro}, J.~F., \& {Steinmetz}, M. 2006, \mnras, 365, 747

\bibitem[{{Agnello} \& {Evans}(2012)}]{agnello12}
{Agnello}, A. \& {Evans}, N.~W. 2012, \mnras, 422, 1767

\bibitem[{{Allende Prieto} {et~al.}(2014){Allende Prieto},
  {Fern{\'a}ndez-Alvar}, {Schlesinger}, {et~al.}}]{allende14}
{Allende Prieto}, C., {Fern{\'a}ndez-Alvar}, E., {Schlesinger}, K.~J., {et~al.}
  2014, \aap, 568, A7

\bibitem[{{Allende Prieto} {et~al.}(2008){Allende Prieto}, {Sivarani}, {Beers},
  {et~al.}}]{allende08}
{Allende Prieto}, C., {Sivarani}, T., {Beers}, T.~C., {et~al.} 2008, \aj, 136,
  2070

\bibitem[{{Beers} {et~al.}(2012){Beers}, {Carollo}, {Ivezi{\'c}},
  {et~al.}}]{beers12}
{Beers}, T.~C., {Carollo}, D., {Ivezi{\'c}}, {\v Z}., {et~al.} 2012, \apj, 746,
  34

\bibitem[{{Bell} {et~al.}(2008){Bell}, {Zucker}, {Belokurov},
  {et~al.}}]{bell08}
{Bell}, E.~F., {Zucker}, D.~B., {Belokurov}, V., {et~al.} 2008, \apj, 680, 295

\bibitem[{{Belokurov} {et~al.}(2014){Belokurov}, {Koposov}, 
  {Evans}, {et~al.}}]{belokurov14}
{Belokurov}, V., {Koposov}, S.~E., {Evans}, N.~W., {et~al.} 2014, \mnras, 437, 116

\bibitem[{Benaglia {et~al.}(2009)Benaglia, Chauveau, Hunter, \&
  Young}]{mixtools09}
Benaglia, T., Chauveau, D., Hunter, D.~R., \& Young, D. 2009, Journal of
  Statistical Software, 32, 1

\bibitem[{{Binney} \& {Merrifield}(1998)}]{binney98}
{Binney}, J. \& {Merrifield}, M. 1998, {Galactic Astronomy} (Princeton
  University Press)

\bibitem[{{Binney} \& {Tremaine}(2008)}]{binney08}
{Binney}, J. \& {Tremaine}, S. 2008, {Galactic Dynamics: Second Edition}
  (Princeton University Press)

\bibitem[{{Bond} {et~al.}(2010){Bond}, {Ivezi{\'c}}, {Sesar},
  {et~al.}}]{bond10}
{Bond}, N.~A., {Ivezi{\'c}}, {\v Z}., {Sesar}, B., {et~al.} 2010, \apj, 716, 1

\bibitem[{{Bovy} {et~al.}(2012){Bovy}, {Allende Prieto}, {Beers},
  {et~al.}}]{bovy12}
{Bovy}, J., {Allende Prieto}, C., {Beers}, T.~C., {et~al.} 2012, \apj, 759, 131

\bibitem[{{Bressan} {et~al.}(2012){Bressan}, {Marigo}, {Girardi},
  {et~al.}}]{bressan12}
{Bressan}, A., {Marigo}, P., {Girardi}, L., {et~al.} 2012, \mnras, 427, 127

\bibitem[{{Brown} {et~al.}(2003){Brown}, {Allende Prieto}, {Beers},
  {et~al.}}]{brown03}
{Brown}, W.~R., {Allende Prieto}, C., {Beers}, T.~C., {et~al.} 2003, \aj, 126,
  1362

\bibitem[{{Brown} {et~al.}(2008){Brown}, {Beers}, {Wilhelm},
  {et~al.}}]{brown08b}
{Brown}, W.~R., {Beers}, T.~C., {Wilhelm}, R., {et~al.} 2008, \aj, 135, 564

\bibitem[{{Brown} {et~al.}(2010){Brown}, {Geller}, {Kenyon}, \&
  {Diaferio}}]{brown10}
{Brown}, W.~R., {Geller}, M.~J., {Kenyon}, S.~J., \& {Diaferio}, A. 2010, \aj,
  139, 59

\bibitem[{{Brown} {et~al.}(2005{\natexlab{a}}){Brown}, {Geller}, {Kenyon}, \&
  {Kurtz}}]{brown05}
{Brown}, W.~R., {Geller}, M.~J., {Kenyon}, S.~J., \& {Kurtz}, M.~J.
  2005{\natexlab{a}}, \apjl, 622, L33

\bibitem[{{Brown} {et~al.}(2005{\natexlab{b}}){Brown}, {Geller}, {Kenyon},
  {et~al.}}]{brown05b}
{Brown}, W.~R., {Geller}, M.~J., {Kenyon}, S.~J., {et~al.} 2005{\natexlab{b}},
  \aj, 130, 1097

\bibitem[{{Bullock} \& {Johnston}(2005)}]{bullock05}
{Bullock}, J.~S. \& {Johnston}, K.~V. 2005, \apj, 635, 931

\bibitem[{{Carollo} {et~al.}(2010){Carollo}, {Beers}, {Chiba},
  {et~al.}}]{carollo10}
{Carollo}, D., {Beers}, T.~C., {Chiba}, M., {et~al.} 2010, \apj, 712, 692

\bibitem[{{Carollo} {et~al.}(2007){Carollo}, {Beers}, {Lee},
  {et~al.}}]{carollo07}
{Carollo}, D., {Beers}, T.~C., {Lee}, Y.~S., {et~al.} 2007, \nat, 450, 1020

\bibitem[{{Chen} {et~al.}(2001){Chen}, {Stoughton}, {Smith}, {et~al.}}]{chen01}
{Chen}, B., {Stoughton}, C., {Smith}, J.~A., {et~al.} 2001, \apj, 553, 184

\bibitem[{{Deason} {et~al.}(2011){Deason}, {Belokurov}, \& {Evans}}]{deason11}
{Deason}, A.~J., {Belokurov}, V., \& {Evans}, N.~W. 2011, \mnras, 416, 2903

\bibitem[{{Deason} {et~al.}(2012){Deason}, {Belokurov}, {Evans}, \&
  {An}}]{deason12}
{Deason}, A.~J., {Belokurov}, V., {Evans}, N.~W., \& {An}, J. 2012, \mnras,
  424, L44

\bibitem[{{Deason} {et~al.}(2013){Deason}, {Van der Marel}, {Guhathakurta},
  {Sohn}, \& {Brown}}]{deason13}
{Deason}, A.~J., {Van der Marel}, R.~P., {Guhathakurta}, P., {Sohn}, S.~T., \&
  {Brown}, T.~M. 2013, \apj, 766, 24

\bibitem[{{Fabricant} {et~al.}(2005){Fabricant}, {Fata}, {Roll},
  {et~al.}}]{fabricant05}
{Fabricant}, D., {Fata}, R., {Roll}, J., {et~al.} 2005, \pasp, 117, 1411

\bibitem[{{Fermani} \& {Sch{\"o}nrich}(2013)}]{fermani13}
{Fermani}, F. \& {Sch{\"o}nrich}, R. 2013, \mnras, 432, 2402

\bibitem[{{Font} {et~al.}(2011){Font}, {McCarthy}, {Crain}, {et~al.}}]{font11}
{Font}, A.~S., {McCarthy}, I.~G., {Crain}, R.~A., {et~al.} 2011, \mnras, 416,
  2802

\bibitem[{{Girardi} {et~al.}(2002){Girardi}, {Bertelli}, {Bressan},
  {et~al.}}]{girardi02}
{Girardi}, L., {Bertelli}, G., {Bressan}, A., {et~al.} 2002, \aap, 391, 195

\bibitem[{{Ibata} {et~al.}(1994){Ibata}, {Gilmore}, \& {Irwin}}]{ibata94}
{Ibata}, R.~A., {Gilmore}, G., \& {Irwin}, M.~J. 1994, \nat, 370, 194

\bibitem[{{Ivezi{\'c}} {et~al.}(2008)}]{ivezic08}
{Ivezi{\'c}}, {\v Z}. {et~al.} 2008, \apj, 684, 287

\bibitem[{{Janesh} {et~al.}(2015){Janesh}, {Morrison}, {Ma},
  {et~al.}}]{janesh15}
{Janesh}, W., {Morrison}, H.~L., {Ma}, Z., {et~al.} 2015, \apj, submitted

\bibitem[{Jeans(1915)}]{jeans15}
Jeans, J.~H. 1915, \mnras, 76, 70

\bibitem[{{Johnston} {et~al.}(2008){Johnston}, {Bullock}, {Sharma}, {Font},
  {Robertson}, \& {Leitner}}]{johnston08}
{Johnston}, K.~V., {Bullock}, J.~S., {Sharma}, S., {et~al.} 2008, \apj, 689, 936

\bibitem[{{Juri\'c} {et~al.}(2008)}]{juric08}
{Juri\'c}, M. {et~al.} 2008, \apj, 673, 864

\bibitem[{{Kafle} {et~al.}(2012){Kafle}, {Sharma}, {Lewis}, \&
  {Bland-Hawthorn}}]{kafle12}
{Kafle}, P.~R., {Sharma}, S., {Lewis}, G.~F., \& {Bland-Hawthorn}, J. 2012,
  \apj, 761, 98

\bibitem[{{Kafle} {et~al.}(2014){Kafle}, {Sharma}, {Lewis}, \&
  {Bland-Hawthorn}}]{kafle14}
---. 2014, \apj, 794, 59

\bibitem[{{King} {et~al.}(2012){King}, {Brown}, {Geller}, \& {Kenyon}}]{king12}
{King}, III, C., {Brown}, W.~R., {Geller}, M.~J., \& {Kenyon}, S.~J. 2012,
  \apj, 750, 81

\bibitem[{{Koposov} {et~al.}(2010){Koposov}, {Rix}, \& {Hogg}}]{koposov10}
{Koposov}, S.~E., {Rix}, H.-W., \& {Hogg}, D.~W. 2010, \apj, 712, 260

\bibitem[{{Kurtz} \& {Mink}(1998)}]{kurtz98}
{Kurtz}, M.~J. \& {Mink}, D.~J. 1998, \pasp, 110, 934

\bibitem[{{Law} \& {Majewski}(2010)}]{law10}
{Law}, D.~R. \& {Majewski}, S.~R. 2010, \apj, 714, 229

\bibitem[{{Lee} {et~al.}(2008){Lee}, {Beers}, {Sivarani}, {et~al.}}]{lee08}
{Lee}, Y.~S., {Beers}, T.~C., {Sivarani}, T., {et~al.} 2008, \aj, 136, 2022

\bibitem[{{Loebman} {et~al.}(2014){Loebman}, {Ivezi{\'c}}, {Quinn},
  {et~al.}}]{loebman14}
{Loebman}, S.~R., {Ivezi{\'c}}, {\v Z}., {Quinn}, T.~R., {et~al.} 2014, \apj,
  794, 151

\bibitem[{{Majewski} {et~al.}(2003){Majewski}, {Skrutskie}, {Weinberg}, \&
  {Ostheimer}}]{majewski03}
{Majewski}, S.~R., {Skrutskie}, M.~F., {Weinberg}, M.~D., \& {Ostheimer}, J.~C.
  2003, \apj, 599, 1082

\bibitem[{{Marigo} {et~al.}(2008){Marigo}, {Girardi}, {Bressan}, {Groenewegen},
  {Silva}, \& {Granato}}]{marigo08}
{Marigo}, P., {Girardi}, L., {Bressan}, A., {et~al.} 2008, \aap, 482, 883

\bibitem[{{McCarthy} {et~al.}(2012){McCarthy}, {Font}, {Crain}, {Deason},
  {Schaye}, \& {Theuns}}]{mccarthy12}
{McCarthy}, I.~G., {Font}, A.~S., {Crain}, R.~A., {et~al.} 2012, \mnras, 420, 2245

\bibitem[{{McMillan} \& {Binney}(2010)}]{mcmillan10}
{McMillan}, P.~J. \& {Binney}, J.~J. 2010, \mnras, 402, 934

\bibitem[{{Mink} {et~al.}(2007){Mink}, {Wyatt}, {Caldwell}, {et~al.}}]{mink07}
{Mink}, D.~J., {Wyatt}, W.~F., {Caldwell}, N., {et~al.} 2007, in Astronomical
  Society of the Pacific Conference Series, Vol. 376, Astronomical Data
  Analysis Software and Systems XVI, ed. R.~A. {Shaw}, F.~{Hill}, \& D.~J.
  {Bell}, 249

\bibitem[{{Monet} {et~al.}(2003){Monet}, {Levine}, {Canzian},
  {et~al.}}]{monet03}
{Monet}, D.~G., {Levine}, S.~E., {Canzian}, B., {et~al.} 2003, \aj, 125, 984

\bibitem[{{Newberg} \& {Yanny}(2006)}]{newberg06}
{Newberg}, H.~J. \& {Yanny}, B. 2006, Journal of Physics Conference Series, 47,
  195

\bibitem[{{Newby} {et~al.}(2011){Newby}, {Newberg}, {Simones}, {Cole}, \&
  {Monaco}}]{newby11}
{Newby}, M., {Newberg}, H.~J., {Simones}, J., {Cole}, N., \& {Monaco}, M. 2011,
  \apj, 743, 187

\bibitem[{{R Core Team}(2014)}]{R14}
{R Core Team}. 2014, R: A Language and Environment for Statistical Computing, R
  Foundation for Statistical Computing, Vienna, Austria

\bibitem[{{Rashkov} {et~al.}(2013){Rashkov}, {Pillepich}, {Deason},
  {et~al.}}]{rashkov13}
{Rashkov}, V., {Pillepich}, A., {Deason}, A.~J., {et~al.} 2013, \apjl, 773, L32

\bibitem[{{Reed}(2006)}]{reed06}
{Reed}, B.~C. 2006, \jrasc, 100, 146

\bibitem[{{Reid} {et~al.}(2014){Reid}, {Menten}, {Brunthaler},
  {et~al.}}]{reid14}
{Reid}, M.~J., {Menten}, K.~M., {Brunthaler}, A., {et~al.} 2014, \apj, 783, 130

\bibitem[{{Reid} {et~al.}(2009){Reid}, {Menten}, {Zheng}, {et~al.}}]{reid09}
{Reid}, M.~J., {Menten}, K.~M., {Zheng}, X.~W., {et~al.} 2009, \apj, 700, 137

\bibitem[{{Schlaufman} {et~al.}(2009){Schlaufman}, {Rockosi}, {Allende Prieto},
  {et~al.}}]{schlaufman09}
{Schlaufman}, K.~C., {Rockosi}, C.~M., {Allende Prieto}, C., {et~al.} 2009,
  \apj, 703, 2177

\bibitem[{{Schlegel} {et~al.}(1998){Schlegel}, {Finkbeiner}, \&
  {Davis}}]{schlegel98}
{Schlegel}, D.~J., {Finkbeiner}, D.~P., \& {Davis}, M. 1998, \apj, 500, 525

\bibitem[{{Sch{\"o}nrich} {et~al.}(2011){Sch{\"o}nrich}, {Asplund}, \&
  {Casagrande}}]{schonrich11}
{Sch{\"o}nrich}, R., {Asplund}, M., \& {Casagrande}, L. 2011, \mnras, 415, 3807

\bibitem[{{Sch{\"o}nrich} {et~al.}(2014){Sch{\"o}nrich}, {Asplund}, \&
  {Casagrande}}]{schonrich14}
---. 2014, \apj, 786, 7

\bibitem[{{Sch{\"o}nrich} {et~al.}(2010){Sch{\"o}nrich}, {Binney}, \&
  {Dehnen}}]{schonrich10}
{Sch{\"o}nrich}, R., {Binney}, J., \& {Dehnen}, W. 2010, \mnras, 403, 1829

\bibitem[{Schwarzschild(1907)}]{schwarzschild1907}
Schwarzschild, K. 1907, Nachrichten von der Gesellschaft der Wissenschaften zu
  G{\"o}ttingen, Mathematisch-Physikalische Klasse, 1907, 614

\bibitem[{{Searle} \& {Zinn}(1978)}]{searle78}
{Searle}, L. \& {Zinn}, R. 1978, \apj, 225, 357

\bibitem[{{Sesar} {et~al.}(2013){Sesar}, {Ivezi{\'c}}, {Stuart},
  {et~al.}}]{sesar13}
{Sesar}, B., {Ivezi{\'c}}, {\v Z}., {Stuart}, J.~S., {et~al.} 2013, \aj, 146,
  21

\bibitem[{{Sesar} {et~al.}(2011){Sesar}, {Juri{\'c}}, \&
  {Ivezi{\'c}}}]{sesar11}
{Sesar}, B., {Juri{\'c}}, M., \& {Ivezi{\'c}}, {\v Z}. 2011, \apj, 731, 4

\bibitem[{{Sirko} {et~al.}(2004){Sirko}, {Goodman}, {Knapp},
  {et~al.}}]{sirko04}
{Sirko}, E., {Goodman}, J., {Knapp}, G.~R., {et~al.} 2004, \aj, 127, 914

\bibitem[{{Smith} {et~al.}(2009){Smith}, {Wyn Evans}, \& {An}}]{smith09a}
{Smith}, M.~C., {Wyn Evans}, N., \& {An}, J.~H. 2009, \apj, 698, 1110

\bibitem[{{Toomet} {et~al.}(2012){Toomet}, {Henningsen}, \& {with contributions
  from Graves, S. and Croissant, Y.}}]{maxlik12}
{Toomet}, O., {Henningsen}, A., \& {with contributions from Graves, S. and
  Croissant, Y.} 2012, maxLik: Maximum Likelihood Estimation, {R package
  version 1.1-2}

\bibitem[{{Wang} {et~al.}(2015){Wang}, {Han}, {Cooper}, {Cole}, {Frenk}, {Cai},
  \& {Lowing}}]{wang15}
{Wang}, W., {Han}, J., {Cooper}, A., {et~al.} 2015, \mnras, 453, 377

\bibitem[{{Watkins} {et~al.}(2009){Watkins}, {Evans}, {Belokurov},
  {et~al.}}]{watkins09}
{Watkins}, L.~L., {Evans}, N.~W., {Belokurov}, V., {et~al.} 2009, \mnras, 398,
  1757

\bibitem[{{Xue} {et~al.}(2011){Xue}, {Rix}, {Yanny}, {et~al.}}]{xue11}
{Xue}, X.-X., {Rix}, H.-W., {Yanny}, B., {et~al.} 2011, \apj, 738, 79

\bibitem[{{Xue} {et~al.}(2008){Xue}, {Rix}, {Zhao}, {et~al.}}]{xue08}
{Xue}, X.~X., {Rix}, H.~W., {Zhao}, G., {et~al.} 2008, \apj, 684, 1143

\bibitem[{{Yanny} {et~al.}(2000){Yanny}, {Newberg}, {Kent}, {et~al.}}]{yanny00}
{Yanny}, B., {Newberg}, H.~J., {Kent}, S., {et~al.} 2000, \apj, 540, 825

\bibitem[{{Zolotov} {et~al.}(2009){Zolotov}, {Willman}, {Brooks},
  {et~al.}}]{zolotov09}
{Zolotov}, A., {Willman}, B., {Brooks}, A.~M., {et~al.} 2009, \apj, 702, 1058

\end{thebibliography}

\clearpage 

\appendix

\section{APPENDIX\label{sec: Appendix}}

\subsection{Coordinate Systems and Projection Factors \label{app: 
CoordinateSystems}}

	We employ four coordinates systems: heliocentric Cartesian $\left( 
x,y,z\right)$ and spherical $\left( d,b,l\right)$ coordinates, and 
Galactocentric Cartesian $\left( X,Y,X\right) $ and spherical $\left( 
R,\theta,\phi\right) $ coordinates. In the Cartesian coordinate systems, the 
Galactocentric $X$- and $Y$-axes lie in the Galactic plane; the 
heliocentric $xy$-plane is parallel to, but slightly above, the Galactic 
plane since the Sun lies slightly above it. The $z$- and $Z$-axes are 
perpendicular to the $xy$- and $XY$-planes, respectively, forming 
right-handed coordinate systems. The positive $x$- and $X$-directions are 
defined as from the Sun toward the Galactic center. Since the Sun is located 
8 kpc from the Galactic center and 0.0196 kpc above the Galactic plane 
\citep{reed06}, the two Cartesian coordinate systems transform as 
\begin{equation} X=x-8.0,\text{ \ \ }Y=y\text{, \ \ }Z=z+0.0196\text{ kpc} 
\end{equation}

	The spherical coordinates are defined in terms of their Cartesian 
coordinates with radii $d=\sqrt{x^{2}+y^{2}+z^{2}}$ and $R=\sqrt{ 
X^{2}+Y^{2}+Z^{2}}$\ for the heliocentric and Galactocentric systems, 
respectively. Longitudes $l$ and $\phi $ are measured in the $xy$- or $XY$- 
plane from the $x$- or $X$-axis toward the $y$- or $Y$-axis. The second angle 
of the spherical coordinates is defined differently for the heliocentric and 
Galactocentric systems. For the heliocentric system, $b$ is defined as the 
latitude, the angle measured from the $xy$-plane with the direction toward 
the $z$-axis taken as positive. For the Galactocentric system, $\theta $ is 
defined as the colatitude, the angle measured from the $Z$-axis.

	The line-of-sight velocity component projection factors are:
\begin{align}
p_{R}& =\cos b\cos l\sin \theta \cos \phi 
  +\cos b\sin l\sin \theta \sin \phi +\sin b\cos \theta \\
p_{\theta }& =\cos b\cos l\cos \theta \cos \phi 
  +\cos b\sin l\cos \theta \sin \phi -\sin b\sin \theta \\
p_{\phi }& =-\cos b\cos l\sin \phi +\cos b\sin l\cos \phi
\end{align}
	A projection factor $p_{\theta }=0.3$, for example, means that a 
tangential velocity component of $v_{\theta }=100$ km~s$^{-1}$ contributes 
$30$ km~s$^{-1}$ to the line-of-sight velocity, $v_{los}$.

\subsection{Probability Density Function \label{app: PDF}}

	Following standard theoretical development, we assume that the 
components of stellar velocity, $v_{i}$, are normally distributed with 
means, $\mu _{i}$, and standard deviations, $\sigma _{i}$,
	\begin{equation} 
v_{i}\sim N\left( \mu _{i},\sigma _{i}^{2}\right) ,\qquad i\in \left\{
R,\theta ,\phi \right\} \text{.}
	\end{equation}
The observed line-of-sight velocity for a given star is then the sum of the 
projections of each velocity component onto the line of sight, as previously 
defined in Equation \ref{eqn: vlos}.  Because the line-of-sight velocity is a 
linear combination of the velocity components, assumed to be normally 
distributed, the observed line-of-sight velocity is also normally distributed 
with mean,
	\begin{equation}
E[v_{los}]=p_{R}\mu _{R}+p_{\theta }\mu _{\theta }+p_{\phi }\mu _{\phi }=%
\mathbf{p}^{\mathbf{\prime }}\boldsymbol{\mu },
	\end{equation}
and variance,
	\begin{equation}
Var[v_{los}]=Var[\mathbf{p}^{\mathbf{\prime }}\mathbf{v]=p}^{\prime }%
\boldsymbol{\Sigma }\mathbf{p},
	\end{equation}
where $\boldsymbol{\Sigma }\mathbf{=}\left[ \Sigma _{ij}\right] $ is the
variance-covariance matrix among the $v_{i}$ and is symmetric, $\Sigma
_{ij}=\Sigma _{ji}$. Combining these results yields the distribution of
line-of-sight velocities
	\begin{align}
v_{los}& \sim N\left( \mathbf{p}^{\mathbf{\prime }}\boldsymbol{\mu }\mathbf{%
,p}^{\prime }\boldsymbol{\Sigma }\mathbf{p}\right)  \\
& =N(p_{R}\mu _{R}+p_{\theta }\mu _{\theta }+p_{\phi }\mu _{\phi }, 
\nonumber \\
& p_{R}^{2}\Sigma _{RR}+p_{\theta }^{2}\Sigma _{\theta \theta }+p_{\phi
}^{2}\Sigma _{\phi \phi }+2p_{R}p_{\theta }\Sigma _{R\theta }+2p_{R}p_{\phi
}\Sigma _{R\phi }+2p_{\theta }p_{\phi }\Sigma _{\theta \phi })  \nonumber
	\end{align}

	Now consider the probability $f\left( x|\boldsymbol{\Theta }\right) $ for
finding a star at Galactic coordinates $\left( R,\theta ,\phi \right) $ with
line-of-sight velocity $v_{los}$. Defining an observation as $x=\left\{
R,\theta ,\phi ,v_{los}\right\} $ and the model parameters as $\boldsymbol{
\Theta }=\left\{ \boldsymbol{\mu }\mathbf{,}\boldsymbol{\Sigma }\right\} $,
the probability distribution function is
	\begin{equation}
f\left( x{|}\boldsymbol{\Theta }\right) =\frac{{\rho \left( \vec{R}\right) }
}{{\sqrt{2\pi }\sqrt{\mathbf{p}^{\prime }\boldsymbol{\Sigma }\mathbf{p}}}}
\exp \left[ {\frac{1}{2}\frac{{{{\left( {{v_{los}}-\mathbf{p}^{
\mathbf{\prime }}\boldsymbol{\mu }}\right) }^{2}}}}{\mathbf{p}
^{\prime }\boldsymbol{\Sigma }\mathbf{p}}}\right] \propto \frac{{1}}{\sqrt{
\mathbf{p}^{\prime }\boldsymbol{\Sigma }\mathbf{p}}}\exp \left[ {\frac{1}{2}
\frac{{{{\left( {{v_{los}}- \mathbf{p}^{\mathbf{\prime }}\boldsymbol{
\mu } }\right) }^{2}}}}{\mathbf{p}^{\prime }\boldsymbol{\Sigma }
\mathbf{p}}}\right] \text{,}
	\end{equation} where $\rho \left( \vec{R}\right) $ is the number 
density of stars. This generalization of the Schwarzschild distribution, 
originally proposed by \citet{schwarzschild1907} to model the velocity 
distribution of stars in the solar neighborhood, further allows correlations 
among the spatial velocities. Since ${\rho \left( \vec{R}\right) }$ is not a 
function of the model parameters, it may be omitted as it does not enter into 
the subsequent maximum likelihood analysis, an advantage of the maximum 
likelihood technique. We estimate nine parameters of $\boldsymbol{\Theta }$ 
in our statistical model:  three means $\mu _{i}$ and six independent 
elements of the symmetric covariance matrix $\boldsymbol{\Sigma }$.

	Denoting an individual observation by $x=\left\{ R,\theta ,\phi 
,v_{los}\right\} $ and a set of observations by $\mathbf{x}=\left\{ 
x_{1},x_{2}...x_{n}\right\} $, the log likelihood function is
	\begin{equation}
\mathcal{L}\left( \mathbf{\Theta }|\mathbf{x}\right)
=\sum_{i=1}^{n}\log f\left( x_{i}|
\mathbf{\Theta }\right) \text{.}
	\end{equation}

\subsection{Data Table\label{app: DataTable}}

	Table \ref{tab: Hecto data} presents the clean sample of 6,174 F-type
stars from the Hectospec radial velocity survey.  For each star we list the
position (epoch J2000), our heliocentric radial velocity measurement, the
SDSS DR10 de-reddened $r$-band magnitude, our estimated absolute magnitude
$M_r$ using the \citet{ivezic08} photometric parallax relation, and the
corresponding Galactocentric radial $R$ and vertical $Z$ distances,
calculated assuming the Sun is at $R=8$ kpc.  The full version of Table
\ref{tab: Hecto data} is available in the online journal.

\begin{deluxetable}{ccrrrrr}            
\tabletypesize{\footnotesize}
\tablewidth{0pt}
\tablecaption{Hectospec F Stars\label{tab: Hecto data}}
\tablecolumns{7}
\tablehead{
  \colhead{R.A.}    & \colhead{Decl.}   & \colhead{$v_{\rm helio}$} & 
  \colhead{$r_0$}   & \colhead{$M_r$}   & \colhead{$R$}             & \colhead{$Z$} \\
  \colhead{(h:m:s)} & \colhead{(d:m:s)} & \colhead{(km s$^{-1}$)}   &
  \colhead{(mag)}   & \colhead{(mag)}   & \colhead{(kpc)}           & \colhead{(kpc)}
}

\startdata
 0:39:30.452 & $+$25:14:52.60 & $-287.6\pm19.8$ & $19.940\pm0.027$ & 5.241 & 13.94 & $ -5.31$ \\
 0:39:59.518 & $+$25:12:41.95 & $ -44.8\pm18.7$ & $20.182\pm0.028$ & 5.821 & 12.89 & $ -4.54$ \\
 0:41:30.298 & $+$25:09:09.98 & $-167.8\pm26.6$ & $20.520\pm0.043$ & 6.412 & 12.25 & $ -4.05$ \\
 0:41:43.272 & $+$25:55:22.42 & $-224.6\pm12.4$ & $19.196\pm0.018$ & 4.790 & 13.07 & $ -4.57$ \\
 0:41:44.054 & $+$25:01:11.58 & $ -75.9\pm23.3$ & $20.510\pm0.037$ & 5.247 & 16.22 & $ -6.92$ \\
\enddata
	\tablecomments{(This table is available in its entirety in 
machine-readable and Virtual Observatory forms in the online journal. A 
portion is shown here for guidance regarding its form and content.)}

\end{deluxetable}


\subsection{Results Table \label{app: ResultsTable}}

Table \ref{tab: Results} presents the estimation results for the 
Hectospec F star and combined samples. 

\begin{deluxetable}{cccccccccccccc}
\tabletypesize{\scriptsize}
\tablewidth{0pt}
\tablecolumns{14}
\tablecaption{Estimation Results \label{tab: Results}}
\tablehead{\colhead{$N$}                               &
           \colhead{$R_{min}$}                         &
           \colhead{$R_{max}$}                         & 
           \colhead{$\bar{R}$}                         &
           \colhead{$\mu_{R}$}                         &
           \colhead{$\mu_{\theta }$}                   &
           \colhead{$\mu_{\phi }$}                     &
           \colhead{$\sigma_{R}$}                      &
           \colhead{$\sigma_{\theta }$}                &
           \colhead{$\sigma_{\phi }$}                  &
           \colhead{$\Sigma_{R \theta}$}               &
           \colhead{$\Sigma_{R \phi}$}                 &
           \colhead{$\Sigma_{\theta \phi}$}            &
           \colhead{$\beta$}                           \\
           \colhead{}                                  &
           \colhead{(kpc)}                             &
           \colhead{(kpc)}                             & 
           \colhead{(kpc)}                             &
           \colhead{(km s$^{-1}$)}                     &
           \colhead{(km s$^{-1}$)}                     & 
           \colhead{(km s$^{-1}$)}                     &
           \colhead{(km s$^{-1}$)}                     &
           \colhead{(km s$^{-1}$)}                     & 
           \colhead{(km s$^{-1}$)}                     &
           \colhead{(km$^{2}$ s$^{-2}$)}               &
           \colhead{(km$^{2}$ s$^{-2}$)}               & 
           \colhead{(km$^{2}$ s$^{-2}$)}               &
           \colhead{} } 
\startdata
\cutinhead{Hectospec F Star Sample}
762 & 6.0   & 10.6  & 9 .3  &     &     &     & 130.3  & 111.5  &  90.0  &     &     &     &  0.40  \\
    &       &       &       &     &     &     & (20.6) & (10.7) & (29.5) &     &     &     & (0.38) \\
762 & 10.6  & 12.4  & 11.4  &     &     &     & 157.5  &  42.5  &  72.8  &     &     &     &  0.86  \\
    &       &       &       &     &     &     & (21.6) & (93.2) & (58.2) &     &     &     & (0.33) \\
762 & 12.4  & 14.9  & 13.5  &     &     &     &  94.7  & 167.0  & 165.6  &     &     &     & -2.08  \\
    &       &       &       &     &     &     & (26.7) & (41.4) & (29.6) &     &     &     & (2.91) \\
763 & 14.9  & 29.8  & 18.2  &     &     &     &  98.9  & 203.3  & 132.3  &     &     &     & -2.01  \\
    &       &       &       &     &     &     &  (9.2) & (30.7) & (35.0) &     &     &     & (1.44) \\
	
\cutinhead{Combined Sample - equally populated bins}
2483  & 6.0   & 9.6   & 8.4   &  0.8  & -4.9  & -7.0  & 155.3  & 109.8  &  88.3  &   428.7  &  -271.8  &   -80.4  &  0.59  \\
      &       &       &       & (9.1) & (5.6) & (7.1) & (14.0) &  (7.8) & (18.6) & (1971.1) & (2251.1) & (1670.5) & (0.13) \\
2482  & 9.6   & 10.6  & 10.1  & -4.6  & -8.0  &  3.9  & 156.5  &  98.2  &  86.2  &  -123.2   & 2442.2  &  -456.0  &  0.65  \\
      &       &       &       & (6.9) & (5.7) & (7.4) & (16.5) & (14.2) & (28.7) & (1949.6) & (2944.3) & (2712.9) & (0.19) \\
2482  & 10.6  & 11.5  & 11.1  & -8.5  & -4.7  & -7.2  &  107.4 &  117.6 & 110.3  & -2806.8  & -2923.5  & -6839.6  & -0.13  \\
      &       &       &       & (5.1) & (5.4) & (7.7) & (21.5) & (20.3) & (26.3) & (1190.2) & (2357.3) & (2723.3) & (0.83) \\
2482  & 11.5  & 12.5  & 12.0  &  4.9  &  3.5  &  3.8  & 150.1 &   38.5  &  85.9  & -2088.1  &   530.2  & -4335.6  &  0.80  \\
      &       &       &       & (4.6) & (5.7) & (9.0) & (12.3) & (76.6) & (40.2) & (1071.4) & (2054.6) & (2761.1) & (0.29) \\
2483  & 12.5  & 13.7  & 13.1  & -8.1  & -1.2  & -8.6  & 105.0  & 165.4  & 194.0  & 1896.4   &  5801.2  &  5498.6  & -1.95  \\
      &       &       &       & (3.9) & (6.1) & (9.7) & (11.8) & (17.8) & (17.9) & (913.9)  & (1961.8) & (3269.4) & (1.11) \\
2482  & 13.7  & 15.3  & 14.4  & -8.3  &  6.3  & -12.7 &  95.1  & 205.3  & 172.6  & 3053.6   &  1448.5  &  2843.3  & -2.97  \\
      &       &       &       & (3.6) & (7.3) & (9.4) & (10.6) & (18.3) & (19.2) & (1023.6) & (1732.2) & (4172.5) & (1.52) \\
2482  & 15.3  & 18.5  & 16.7  & -8.6  & -19.4 & -40.1 &  56.8  & 256.0  & 225.8 & 1243.5 & 494.8 & 831.7 & -17.08          \\
      &       &       &       & (2.9) & (8.3) & (9.5) &  (9.3) & (14.4) & (14.2) & (793.6) & (1257.0) & (4391.9) & (7.50)  \\
2483  & 18.5  & 30.0  & 22.4  & -10.7 & -24.8 & -56.6 &  76.8  & 259.1  & 195.8 & 57.2  & 30.9  & 10559.0 & -7.95          \\
      &       &       &       & (2.4) & (11.3) & (10.1) & (4.7) & (18.8) & (21.7) & (952.3) & (1046.3) & (6406.6) & (2.32) \\

\cutinhead{Combined Sample - $4$ kpc bins}
    3356  & 6     & 10    & 8.8   & -7.4  & -8.6  & -9.9  & 149.4 & 109.4 & 91.5  & 65.0  & -438.2 & -212.8 & 0.54 \\
          &       &       &       & (7.3) & (4.8) & (6.1) & (11.6) & (6.6) & (15.8) & (1566.3) & (1951.8) & (1512.3) & (0.12) \\
    5680  & 7     & 11    & 9.7   & -3.6  & -6.3  & -4.7  & 137.7 & 109.8 & 110.5 & -505.9 & 1309.8 & -1025.6 & 0.36 \\
          &       &       &       & (4.6) & (3.6) & (4.7) & (7.5) & (5.3) & (11.0) & (987.4) & (1460.6) & (1263.4) & (0.12) \\
    7915  & 8     & 12    & 10.4  & -2.3  & -4.1  & -2.9  & 129.6 & 109.9 & 112.6 & -1373.8 & 447.5 & -2247.6 & 0.26 \\
          &       &       &       & (3.3) & (3.0) & (4.2) & (5.2) & (4.8) & (9.4) & (692.5) & (1124.0) & (1088.7) & (0.11) \\
    9505  & 9     & 13    & 11.1  & -2.8  & -1.8  & -4.2  & 128.6 & 114.8 & 123.9 & -603.9 & 1515.5 & -1374.5 & 0.14 \\
          &       &       &       & (2.7) & (2.8) & (4.0) & (3.9) & (4.9) & (8.6) & (594.3) & (1023.9) & (1112.6) & (0.11) \\
    9658  & 10    & 14    & 11.9  & -3.8  & -2.4  & -2.1  & 129.3 & 112.7 & 125.0 & -426.5 & 1837.7 & -1941.9 & 0.15 \\
          &       &       &       & (2.3) & (2.8) & (4.2) & (3.5) & (6.2) & (9.5) & (518.2) & (955.3) & (1186.2) & (0.12) \\
    8606  & 11    & 15    & 12.8  & -6.2  & -2.2  & -6.3  & 123.5 & 126.3 & 135.3 & 315.9 & 1725.8 & -471.1 & -0.12 \\
          &       &       &       & (2.2) & (3.2) & (4.8) & (3.9) & (8.2) & (11.0) & (533.1) & (989.9) & (1543.6) & (0.19) \\
    6933  & 12    & 16    & 13.7  & -6.1  & 3.8   & -11.8 & 110.4 & 157.7 & 166.7 & 1687.6 & 2648.4 & 1922.0 & -1.16 \\
          &       &       &       & (2.2) & (3.9) & (5.6) & (4.8) & (9.6) & (11.0) & (578.4) & (1082.8) & (2092.4) & (0.39) \\
    5354  & 13    & 17    & 14.6  & -6.2  & -0.7  & -12.3 & 97.2  & 190.1 & 180.4 & 2195.8 & 2557.4 & 3649.3 & -2.64 \\
          &       &       &       & (2.3) & (4.8) & (6.5) & (5.8) & (11.1) & (12.4) & (641.7) & (1142.6) & (2718.3) & (0.79) \\
    4077  & 14    & 18    & 15.6  & -5.9  & -3.1  & -21.1 & 72.0  & 233.9 & 201.0 & 1907.1 & 1291.7 & 2263.5 & -8.16 \\
          &       &       &       & (2.4) & (6.0) & (7.4) & (6.9) & (10.7) & (12.3) & (696.4) & (1125.7) & (3316.7) & (2.46) \\
    3069  & 15    & 19    & 16.7  & -8.3  & -10.4 & -39.6 & 56.5  & 257.9 & 215.8 & 1672.3 & -556.6 & -1410.4 & -16.71 \\
          &       &       &       & (2.6) & (7.5) & (8.4) & (8.3) & (12.5) & (12.9) & (711.3) & (1107.7) & (3941.0) & (6.56) \\
    2504  & 16    & 20    & 17.7  & -13.3 & -29.2 & -50.8 & 53.1  & 270.2 & 217.6 & 1050.0 & -1129.9 & -84.7 & -20.35 \\
          &       &       &       & (2.6) & (8.9) & (9.2) & (7.2) & (12.8) & (13.4) & (750.1) & (1058.7) & (4577.8) & (7.40) \\
    2013  & 17    & 21    & 18.7  & -12.7 & -25.2 & -59.2 & 58.5  & 257.8 & 218.5 & 334.5 & -1897.6 & -1140.2 & -15.68 \\
          &       &       &       & (2.8) & (10.2) & (10.2) & (6.8) & (16.5) & (15.7) & (786.5) & (1093.2) & (5339.4) & (5.60) \\
    1615  & 18    & 22    & 19.7  & -15.7 & -23.7 & -63.7 & 58.9  & 284.7 & 217.0 & 235.7 & -1582.4 & 4292.7 & -17.48 \\
          &       &       &       & (3.1) & (12.7) & (11.7) & (7.4) & (19.3) & (19.4) & (986.6) & (1276.8) & (7048.2) & (6.75) \\
    1328  & 19    & 23    & 20.7  & -14.3 & -37.3 & -65.4 & 52.6  & 298.9 & 243.7 & -1262.7 & -125.2 & 10052.2 & -25.89 \\
          &       &       &       & (3.2) & (14.3) & (13.4) & (8.5) & (21.4) & (21.6) & (1063.2) & (1321.9) & (7893.0) & (11.88) \\
    1093  & 20    & 24    & 21.8  & -8.9  & -12.4 & -62.1 & 54.4  & 323.9 & 241.3 & 66.4  & -920.7 & 3538.6 & -26.59 \\
          &       &       &       & (3.6) & (17.1) & (15.2) & (9.9) & (26.8) & (25.2) & (1355.4) & (1431.5) & (8801.9) & (13.94)   \\
    918   & 21    & 25    & 22.8  & -9.9  & -26.5 & -75.5 & 48.7  & 362.1 & 248.9 & 1689.7 & 462.6 & 13621.3 & -39.73 \\
          &       &       &       & (4.0) & (20.2) & (16.7) & (11.8) & (28.1) & (26.8) & (1763.6) & (1537.5) & (10629.8) & (25.11) \\
    763   & 22    & 26    & 23.7  & -2.2  & -21.5 & -48.2 & 62.8  & 329.5 & 232.4 & 959.3 & 550.4 & 7405.9 & -19.61 \\
          &       &       &       & (4.3) & (22.2) & (18.4) & (10.4) & (31.7) & (34.1) & (1954.0) & (1713.7) & (12355.0) & (10.39) \\
    649   & 23    & 27    & 24.8  & -0.9  & -12.6 & -50.3 & 85.3  & 279.1 & 142.4 & 1262.5 & 873.7 & 19147.5 & -5.75 \\
          &       &       &       & (4.8) & (25.4) & (20.7) & (10.1) & (46.1) & (70.0) & (2318.5) & (1984.0) & (15005.4) & (4.26) \\
    538   & 24    & 28    & 25.8  & -1.5  & -46.9 & -28.4 & 92.6  & 191.3 & 123.6 & -1807.9 & 2175.8 & 26741.0 & -2.02 \\
          &       &       &       & (5.0) & (26.9) & (23.0) & (11.4) & (74.5) & (112.0) & (2260.6) & (2252.1) & (16965.0) & (3.70) \\
    443   & 25    & 29    & 26.7  & -6.1  & -36.1 & -11.1 & 90.4  & 140.5 & 160.5 & -5339.9 & -251.5 & -2283.8 & -1.78 \\
          &       &       &       & (5.5) & (30.3) & (28.0) & (14.1) & (138.9) & (114.7) & (2179.6) & (2674.5) & (19133.2) & (5.12) 		  
\enddata

\end{deluxetable}

\end{document}